\documentclass[12pt]{article}
\usepackage[dvips]{graphics}
\usepackage{natbib}
\usepackage{epsfig}
\usepackage{color}
\usepackage{lscape}
\usepackage{amsmath}
\usepackage[normalem]{ulem}
\usepackage{tabularx}
\usepackage{mathtools}
\usepackage{enumerate} 
\usepackage{fullpage, amssymb,amsthm,chngcntr}
\usepackage{amsfonts,color}
\usepackage{bm}
\usepackage{hyperref}
\usepackage{verbatim}
\usepackage{epstopdf}
\usepackage{multirow}
\usepackage{setspace}
\usepackage{float}
\usepackage{caption}
\usepackage{subcaption}
\usepackage{booktabs}
\allowdisplaybreaks
\usepackage{enumitem}
\usepackage{appendix}
\usepackage[margin=1in]{geometry}
\usepackage{rotating}
\usepackage{float}
\usepackage{tabularx}
\usepackage{changepage}
\usepackage{siunitx} 
\usepackage{makecell} 
\usepackage{graphicx} 
\usepackage{boldline} 
\usepackage[utf8]{inputenc}
\usepackage{algpseudocode}
\usepackage{algorithm}
\usepackage{color}
\usepackage[svgnames]{xcolor}

\newcommand{\va}{{\bf a}}

\newcommand{\vx}{{\bf x}}

\newcommand{\vnull}{{\bf 0}}

\newcommand{\vSigma}{\mbox{\boldmath $\Sigma$}}

\newcommand{\vbeta}{\mbox{\boldmath $\beta$}}

\newcommand{\ep}{\epsilon}

\newcommand{\bay}{\begin{array}}
\newcommand{\eay}{\end{array}}

\newcommand{\bqa}{\begin{eqnarray*}}
\newcommand{\eqa}{\end{eqnarray*}}
\newcommand{\bqan}{\begin{eqnarray}}
\newcommand{\eqan}{\end{eqnarray}}
\newcommand{\bqt}{\begin{quote}}
\newcommand{\eqt}{\end{quote}}
\newcommand{\bt}{\begin{tabbing}}
\newcommand{\et}{\end{tabbing}}
\newcommand{\bit}{\begin{itemize}}
\newcommand{\eit}{\end{itemize}}
\newcommand{\ben}{\begin{enumerate}}
\newcommand{\een}{\end{enumerate}}
\newcommand{\beq}{\begin{equation}}
\newcommand{\eeq}{\end{equation}}
\newcommand{\bdefi}{\begin{definition}}
\newcommand{\edefi}{\end{definition}}
\newcommand{\bpro}{\begin{proposition}}
\newcommand{\epro}{\end{proposition}}
\newcommand{\bco}{\begin{corollary}}
\newcommand{\eco}{\end{corollary}}
\newcommand{\bdes}{\begin{description}}
\newcommand{\edes}{\end{description}}

\def\wh{\widehat}

\def\log{\hbox{log}}

\def\boxit#1{\vbox{\hrule\hbox{\vrule\kern6pt
          \vbox{\kern6pt#1\kern6pt}\kern6pt\vrule}\hrule}}

\def\bse{\begin{eqnarray*}}
\def\ese{\end{eqnarray*}}
\def\be{\begin{eqnarray}}
\def\ee{\end{eqnarray}}
\def\bq{\begin{equation}}
\def\eq{\end{equation}}

\def\wh{\widehat}

\def\trans{^{\top}}

\newtheorem{proposition}{Proposition}

\newcommand{\blem}{\begin{lemma}}
\newcommand{\elem}{\end{lemma}}
\newcommand{\bthe}{\begin{theorem}}
\newcommand{\ethe}{\end{theorem}}

\newtheorem{definition}{Definition}[section]
\newtheorem{lemma}[definition]{Lemma}
\newtheorem{theorem}[definition]{Theorem}

\def\delete#1{\iffalse #1 \fi}

\def\bse{\begin{eqnarray*}}
\def\ese{\end{eqnarray*}}
\def\bee{\begin{enumerate}}
\def\eee{\end{enumerate}}
\def\bqe{\begin{eqnarray}}
\def\eqe{\end{eqnarray}}
\def\bed{\begin{description}}
\def\eed{\end{description}}
\def\bei{\begin{itemize}}
\def\eei{\end{itemize}}

\def\pmb#1{\setbox0=\hbox{#1}%
    \kern-.025em\copy0\kern-\wd0
    \kern.05em\copy0\kern-\wd0
    \kern-.025em\raise.0433em\box0 }
\def\pmbh#1#2{\setbox0=\hbox{#1}%
    \setbox1=\hbox{#2}%
    \kern-.025em\copy0\kern-\wd0
    \kern.05em\copy1\kern-\wd0
    \kern-.025em\raise.0433em\box0 }

\def\frac#1#2{{#1\over#2}}

\def\boxit#1{\vbox{\hrule\hbox{\vrule\kern6pt
   \vbox{\kern6pt#1\kern6pt}\kern6pt\vrule}\hrule}}

\def\listing#1{\vskip 4mm\begin{verbatim}\input#1 \vskip 4mm}
\def\thick#1{\hbox{\rlap{$#1$}\kern0.25pt\rlap{$#1$}\kern0.25pt$#1$}}

\def\wh{\widehat}

\def\pmbh{{\pmb h}}

\renewcommand\today{\ifcase\month\or
   Jan\or Feb\or Mar\or Apr\or May\or
   Jun\or Jul\or Aug\or Sep\or Oct\or Nov\or
   Dec\fi
   \space\number\day, \number\year}

\newtheorem{thm}{Theorem}
\newtheorem{lem}{Lemma}

\newtheorem{rmk}{Remark}

\usepackage{lipsum}
\makeatletter
\def\ps@pprintTitle{%
	\let\@oddhead\@empty
	\let\@evenhead\@empty
	\def\@oddfoot{}%
	\let\@evenfoot\@oddfoot}
\makeatother
\DeclareMathOperator*{\argmin}{arg\,min}

\begin{document}
\renewcommand{\thepage}{}
\begin{singlespace}
\title{ \bf Identifying Genetic Variants for Obesity: A Knowledge Integration Quantile Regression (KIQR) Approach for Ultra-High-Dimensional Data  \vspace{-2em}}
\date{}
\maketitle
\begin{center}
    \textbf{Jiantong Wang$^{a}$, Heng Lian$^{b}$, Yan Yu$^{c}$, Tianhai Zu$^{d}$, and Heping Zhang$^{e}$}
\end{center} 		       

\begin{center}
	$^{a}$Department of Business Analytics \& Information Systems, Xavier University, \\Cincinnati, OH \\
    $^{b}$Department of Mathematics, City University of Hong Kong, Kowloon, Hong Kong\\
    $^{c}$Department of Operations, Business Analytics, \& Information Systems,\\ University of Cincinnati, Cincinnati, OH \\
    $^{d}$Department of Management Science and Statistics, University of Texas at San Antonio, San Antonio, TX\\
    $^{e}$Department of Biostatistics, Yale University, New Haven, CT\\
\end{center}

\begin{abstract}
Obesity is widely recognized as a serious and pervasive health concern. We study obesity through body mass index (BMI), which is known to be highly heritable, and
identify important genetic risk factors for BMI from hundreds of thousands of single nucleotide polymorphisms (SNPs) in the Framingham Study data. Several challenges arise when using traditional genome-wide association studies (GWAS): (1) They suffer from a low power due to a combination of a limited number of participants and the stringent genome-wide significance threshold; (2) existing prior knowledge from large meta-analyses may provide valuable guidance but is often underutilized; (3) the one-at-a-time univariate marginal regression framework ignores the joint and conditional nature of genetic effects; (4) GWAS focus solely on mean outcomes, whereas obesity inherently concerns abnormally high BMI levels. To address these challenges, we conduct the analysis by proposing and applying a novel Knowledge Integration Quantile Regression (KIQR) approach via simultaneous variable selection and estimation, focusing on the conditional high quantiles of BMI, which are most relevant to obesity risk, while integrating prior information from large-scale studies such as the GIANT consortium and UK Biobank. Notably, we identified promising novel associations: rs3798696 in \textit{TFAP2A}, rs7070523 in \textit{ITIH5}, and rs178260 in \textit{AIFM3}, which have not previously been reported in the GWAS literature. These findings provide new insights into the genetic architecture of obesity and demonstrate that quantile-based modeling with integrated prior knowledge can potentially uncover novel genes missed by traditional GWAS approaches. An R implementation and simulation scripts are available at: \url{https://github.com/KIQR-submission/KIQR}.

\end{abstract}
        
\begin{keywords}
    {BMI; GWAS; Penalized regression; SCAD; Ultra-high dimensional data; Variable selection} 
\end{keywords}
    
\end{singlespace}
\clearpage
\pagestyle{plain}
\pagenumbering{arabic}\label{key}
		
\section{Introduction}
The global epidemic of obesity poses a severe threat to public health, compromising individual well-being and placing an overwhelming burden on healthcare systems. According to recent World Health Organization (WHO) reports, over one billion people worldwide are currently classified as obese, and the figure continues to rise rapidly. The human and economic tolls are staggering: obesity contributes to an estimated four million deaths annually, and the 2023 World Obesity Atlas projects that related global costs will reach \$4 trillion annually by 2035.\footnote{https://www.worldobesity.org/resources/resource-library/world-obesity-atlas-2023} It is critical to investigate the underlying risk factors associated with obesity so that effective strategies and interventions can be developed. Despite notable progress in research and treatment strategies, many unknown risk factors remain to be discovered, partly due to the complex nature of obesity \citep{Loos2022}.

When attempting to identify important genetic risk factors associated with obesity by conducting traditional genome-wide association studies (GWAS)\footnote{\cite{Myocardial_Infarction_Genetics_Consortium2009}} using Framingham Heart Study data (FHS), several challenges emerge. The data we use contain fewer than 2,000 participants, yet encompass more than 500,000 single nucleotide polymorphisms (SNPs), posing significant difficulties for traditional GWAS analysis. Standard GWAS procedures evaluate each SNP individually in a univariate manner and apply a stringent genome-wide significance threshold ($5\times 10^{-8}$). The combination of a limited number of participants and stringent significance criteria renders the identification of meaningful variants nearly impossible; as shown in Figure~\ref{fig:GWAS}, no SNPs are identified with traditional GWAS. Even when biologically relevant SNPs exist, they rarely reach the significance threshold required by GWAS.

\begin{figure}[h!]
    \centering
    \caption{Manhattan plot comparing traditional GWAS and the KIQR approach. The horizontal red line represents the genome-wide significance threshold ($P = 5\times10^{-8}$), which no SNPs reach using traditional GWAS. Triangles with text labels indicate a selected subset of novel discoveries.}
    \includegraphics[width=0.95\linewidth]{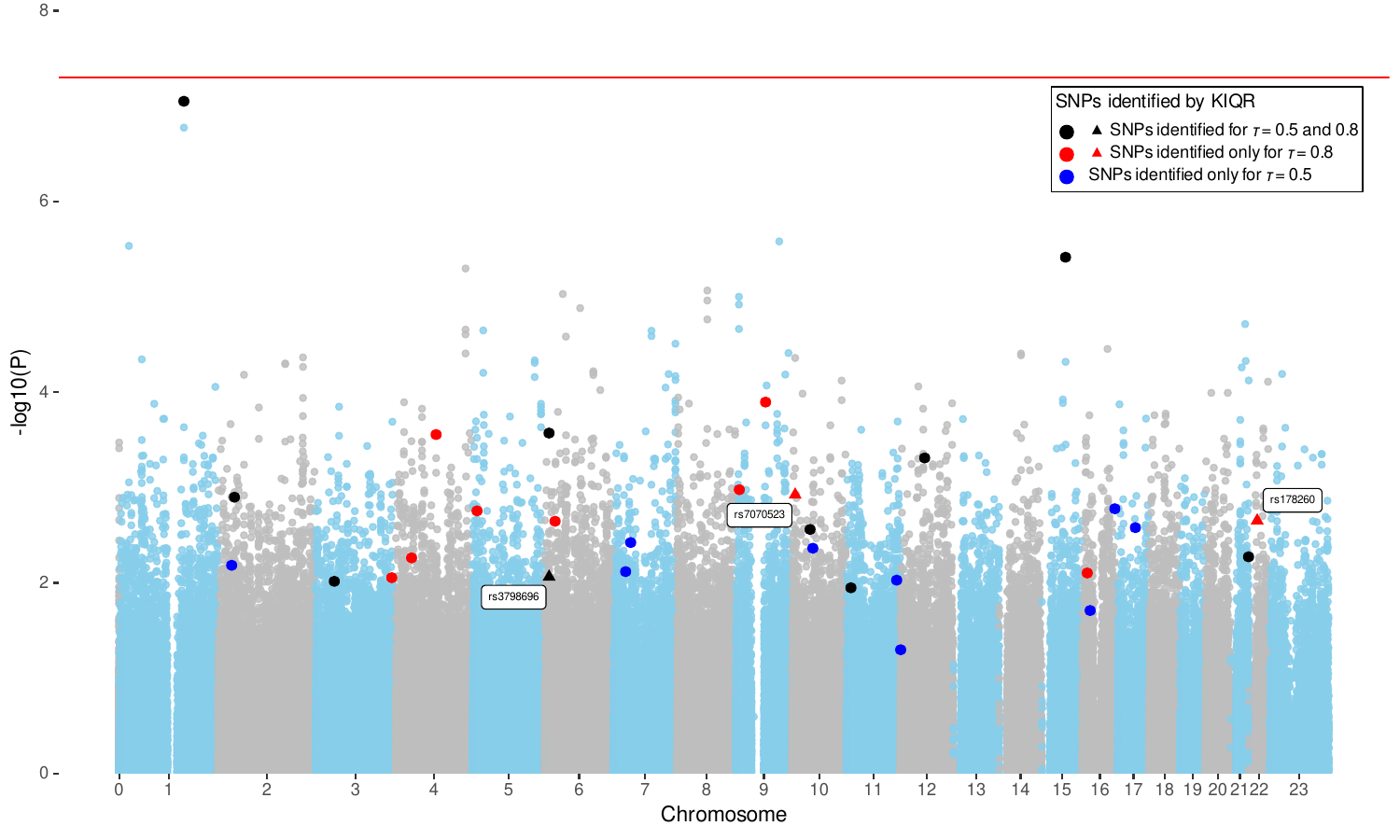}
    \label{fig:GWAS}
\end{figure}

Meanwhile, a substantial body of genetic association findings has emerged from meta-analyses of large-scale studies, including Genetic Investigation of ANthropometric Traits (GIANT) and UK Biobank. Indeed, the majority of currently recognized genetic risk variants are products of extensive meta-analyses of GWAS \citep{Evangelou2013}. However, in cohort studies such as the Framingham Heart Study, such potentially valuable prior knowledge is rarely fully utilized. Integrating prior knowledge is often complicated by differences in study design, phenotype definitions, and population structure across studies. The inherent constraints of standard meta-analysis, especially its requirement for homogeneity in study design and analytical methods \citep{Haidich2010}, highlight the need for more versatile approaches.


The limitations of traditional GWAS extend beyond statistical power to its modeling assumptions, examining each SNP one-at-a-time through marginal regression, thus failing to account for the potential joint and conditional effects among multiple genetic variants that can collectively be associated with obesity. Moreover, standard GWAS focuses solely on genetic associations with respect to the mean level of body mass index (BMI); here BMI is calculated by dividing an individual's weight in kilograms by the square of their height in meters. However, obesity inherently concerns abnormally high BMI levels. As defined by WHO, obesity is a condition characterized by a BMI of 30 or higher. The BMI distribution is also inherently heterogeneous. For obesity research, it is therefore of greater clinical interest to investigate genetic risk factors for high or abnormal quantiles of BMI rather than the mean. Quantile regression, which has shown great success \citep{Koenker1978, Yu2003}, is naturally more suitable for characterizing these conditional high quantiles of BMI that tend to be heterogeneous. Studies focusing on genetic risk factors for obesity using quantile regression remain scarce, particularly in the context of ultra-high-dimensional SNPs.\footnote{ Research on the quantiles of BMI has been recognized by previous researchers investigating fixed phenotype risk factors (e.g., \citealt{Azagba2012, Li2010, Bottai2014}).} In this work, we present the first application of penalized quantile regression with simultaneous variable selection to the genetic study of obesity, focusing specifically on high conditional quantiles of BMI.  

To address these challenges and incorporate valuable prior knowledge, we propose and apply a novel Knowledge Integration Quantile Regression (KIQR) approach that performs simultaneous variable selection and estimation. As illustrated in 
Figure~\ref{fig:workflow}, KIQR operates in two steps: Step 1 constructs a prior-informed estimator via penalized quantile regression, incorporating prior SNP sets derived from large-scale meta-analyses such as the GIANT Consortium and the UK Biobank; Step 2 applies KIQR estimation using Huber loss with an 
LLA-SCAD (the local linear approximation of the SCAD penalty; \citealt{Fan2001, Zou2008}) penalty optimized via cyclic coordinate descent. KIQR is specifically designed to model the conditional high quantiles of BMI most relevant to obesity risk, while maintaining robustness to potential misspecification in prior information. 

As shown in Figure~\ref{fig:GWAS}, although no SNPs reach the 
genome-wide significance threshold under traditional GWAS, KIQR 
identifies novel SNPs at high conditional quantiles of BMI, 
including rs3798696 in \textit{TFAP2A}, rs7070523 in \textit{ITIH5}, 
and rs178260 in \textit{AIFM3}. Furthermore, KIQR recovers the SNPs 
with the strongest marginal association signals 
(e.g., rs2456899 on chromosome~1 and rs10519076 on chromosome~15), 
confirming its validity. A detailed presentation of the KIQR 
methodology and these novel findings is provided in 
Sections~\ref{sec:KIQR} and~\ref{sec:realdata}.

\begin{figure}[h!]
    \centering
    \caption{Overview of the Knowledge Integration Quantile Regression (KIQR) framework. Cohort study data with over half-million SNPs are analyzed using a two-step procedure: Step 1 constructs a prior-informed estimator $\hat{\beta}_p$ via penalized quantile regression, incorporating prior SNP sets $S_p$ derived from large-scale meta-analyses (e.g. UK Biobank \& GIANT Consortium); Step 2 applies KIQR estimation using Huber loss with LLA-SCAD penalty optimized via cyclic coordinate descent, yielding selected SNPs at various quantiles including novel loci \textit{TFAP2A}, \textit{ITIH5}, and \textit{AIFM3}.}
    \includegraphics[width=0.95\linewidth]{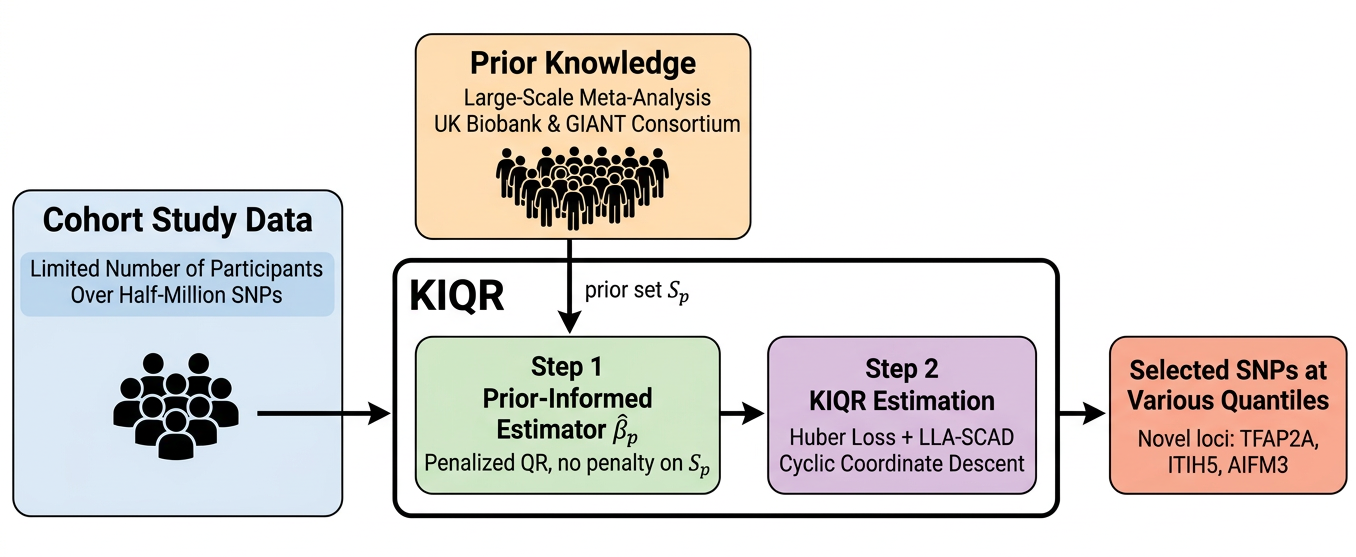}
    \label{fig:workflow}
\end{figure}

To the best of our knowledge, in application, this is the first attempt to simultaneously discover and estimate important genetic risk factors across hundreds of thousands of SNPs for obesity, focusing on the high conditional quantiles of BMI, and flexibly incorporating knowledge from previous studies.

The structure of this article is as follows: Section~\ref{sec:KIQR} introduces the KIQR approach, the approximations introduced for computational efficiency, and the computational algorithm. Section~\ref{sec:realdata} presents our application to obesity analysis and describes newly identified SNPs associated with high quantiles of BMI along with a mimic genetic data simulation study. In Section~\ref{sec:sim}, we present additional numerical simulations. In Section~\ref{sec:theory}, we establish challenging theoretical properties of KIQR in ultra-high-dimensional settings. The proofs are relegated to the Online Supplements. 

\section{Knowledge Integration Quantile Regression (KIQR)}
\label{sec:KIQR}

\subsection{Knowledge Integration Quantile Regression Model for Obesity}  
To identify important SNPs in our analysis, we develop a novel KIQR approach designed for simultaneous variable selection and quantile regression for ultra-high-dimensional data, incorporating knowledge from previous studies in flexible forms. The KIQR model is partly inspired by the pLASSO model from \cite{Jiang2016}, which effectively integrates prior information into the LASSO \citep{Tibshirani1996} framework in the mean regression context. However, the set of genetic variables in their real data analysis is on a much smaller scale, with a total of 916 SNPs for bipolar disorder. More importantly, while quantile regression is desirable, major challenges arise both in theory and in computation because of the non-differentiable check loss function. Consequently, the KIQR solution lacks a closed-form analytical expression and cannot rely on standard LASSO formulations. Instead, to enhance computational efficiency, we employ a Huber-loss approximation (HLA) for the quantile check loss. We adopt the Smooth Clipped Absolute Deviation (SCAD) penalty function \citep{Fan2001} for its desirable oracle properties. We use local linear approximation (LLA) to tackle the difficulties introduced by the nonconvex nature of the SCAD penalty. In theory, we establish challenging asymptotic properties including the oracle properties with these approximations for KIQR in the ultra-high-dimensional setting.

In our obesity analysis, the response variable is BMI, with $\mathbf{y} = (y_1 , y_2 , \dots , y_n)$ denoting the BMI measurements of independent participants $i = 1, \dots , n$. We denote the covariate matrix as $\mathbf{X}_{n \times d}$, which includes genetic and other risk factors, where $\mathbf{x}_i\trans = (x_{i1}, x_{i2}, \dots, x_{id})$ represents the covariate vector of participant $i$. Note that the dimension of the covariates, $d$, is allowed to diverge and can even reach ultra-high dimensionality at an exponential rate of $n$ \citep{FanandLv2008}. We utilize a subset of a popular ongoing NIH study, Framingham Heart Study (FHS) \citep{FHS} with $n = 1,964$ participants. The number of SNPs analyzed exceeds 500,000, which is much higher than the number of participants.  
		
Fortunately, there are many well-established studies investigating the genetic risk factors of obesity, albeit across cohorts. Notably, some of these investigations have been conducted using large biobank studies, such as the UK Biobank. Furthermore, international collaborations, such as the Genetic Investigation of ANthropometric Traits (GIANT) consortium, have played a vital role in advancing our understanding of obesity. The GIANT consortium, which involves a collaboration of researchers from various institutions, countries, and studies, has primarily focused on the meta-analysis of genome-wide association data and other extensive genetic datasets \citep{Yengo2018}. Based on these foundational works,  we propose to utilize the valuable knowledge derived from these previous studies, especially focusing on the confirmed significant SNPs in earlier meta-analyses.
		
To simultaneously select and estimate important risk factors for high quantiles of BMI with ultra-high-dimensional data, while integrating knowledge from established studies, we propose and employ the KIQR method. For a given quantile level $\tau \in (0,1)$, we employ the conditional quantile regression framework defined as follows to estimate and select variables from ultra-high-dimensional covariates,
    \begin{align}
		\theta_{\tau}(\mathbf{y}_i |  \mathbf{x}_i ) = \mathbf{x}_i\trans \boldsymbol{\beta}_{\tau}. \label{eq:conditional_quantile}
    \end{align}
where $\boldsymbol{\beta}_{\tau} = (\beta_{\tau 0}, \beta_{\tau 1}, \beta_{\tau 2}, \dots, \beta_{\tau d})\trans$. In our setting, $\boldsymbol{\beta}_{\tau}$ can vary across different conditional quantile levels $\tau$, which allows for the selection of different genetic risk factors associated with different conditional quantiles of BMI. 
We omit the subscript $\tau$ for notational simplicity and denote the coefficient vector as $\boldsymbol{\beta} = (\beta_0, \beta_1, \beta_2, \dots, \beta_{d})\trans$. 
		
To achieve simultaneous variable selection and modeling, we employ the framework of penalized quantile regression proposed by \cite{wuliu09}, whose objective function is:
    \begin{align*}
    \mathbf{Q_{\lambda}} (\boldsymbol{\beta} ; \mathbf{X}, \mathbf{y}) &= \frac{1}{n} \sum_{i=1}^{n} \rho_{\tau}(y_i - \mathbf{x}_i{\trans} \boldsymbol{\beta}) + \sum_{j = 1}^{d}p_{\lambda}(|{\beta_j}|),
    \end{align*}
where $ \rho_{\tau}(u) = u \{ \tau - \mathbf{I} (u < 0) \} $ is the quantile check loss function, and $p_{\lambda}(\cdot)$ is the penalty function with tuning parameter $\lambda$. While various popular penalty functions can be employed, we specifically utilize the Smooth Clipped Absolute Deviation (SCAD) penalty function in our application for its appealing oracle property. The SCAD penalty is a nonconvex function defined by $p_{\lambda}(0) = 0$ and $p^{\prime}_{\lambda}(\beta)=\lambda\left[I(\beta \leq \lambda)+\frac{(a \lambda-\beta)_{+}}{(a-1) \lambda} I(\beta>\lambda)\right]$ for $|\beta| > 0$, where $\lambda$ is a regularization penalty parameter. We adopt $a = 3.7$ as recommended by \cite{Fan2001}.

Our approach provides the flexibility to accommodate diverse knowledge formats by integrating information via a single prior prediction vector $\hat{y}_i^p$. For instance, known predicted effect sizes $\hat{\boldsymbol{\beta}}^p$ established in a previous study can be used to directly compute the prediction vector $\hat{y}_i^p = \mathbf{x}_i^\top \hat{\boldsymbol{\beta}}^p$. Similarly, an established model from a previous study could be used to generate the prediction vector $\hat{y}_i^p$ directly. In our application, we utilize the most common format of prior knowledge, important variables identified in earlier research, which we denote as the prior set $\mathbf{S}_p$, as shown in Figure~\ref{fig:workflow}. This format requires Step 1 to obtain the prior-informed estimator $\hat{\boldsymbol{\beta}}^p$ by fitting a traditional penalized quantile regression model, where the variables in the prior set $\mathbf{S}_p$ are unpenalized:
    \begin{align}
    \hat{\boldsymbol{\beta}}^p &= \underset{\boldsymbol{\beta}}{\text{arg min}}	\{\mathbf{Q}_{\lambda,\mathbf{S}_p}(\boldsymbol{\beta}; \mathbf{X}, \mathbf{y})\} = \frac{1}{n} \sum_{i=1}^{n} \rho_{\tau}(y_i - \mathbf{x}_i\trans \boldsymbol{\beta}) + \sum_{j \not\in \mathbf{S}_p}^{}p_{\lambda}(|{\beta_j}|).
    \label{betap}
    \end{align}
		
In our obesity analysis, the prior set $\mathbf{S}_p$ is composed of important SNPs identified from the meta-analysis of large-scale GWAS, including the UK Biobank and GIANT consortium \citep{Yengo2018}. The resulting estimator $\hat{\boldsymbol{\beta}}^p$ from Equation~(\ref{betap}) is then integrated into Step 2 KIQR estimation as the prior prediction vector $\mathbf{x}_i^\top \hat{\boldsymbol{\beta}}^p$. The objective function of KIQR balances the quantile loss from the observed data with the quantile loss from these prior-informed predictions:
    \begin{align}
    \mathbf{Q}_{\lambda, \zeta}(\boldsymbol{\beta}; \mathbf{X}, \mathbf{y}, \hat{\boldsymbol{\beta}}^p) 
    &= \frac{1-\zeta}{n} \sum_{i=1}^{n} \rho_{\tau}(y_i - \mathbf{x}_i\trans \boldsymbol{\beta}) 
    + (1 - \zeta)\sum_{j = 1}^{d}p_{\lambda}(|\beta_j|) 
    + \frac{\zeta}{n} \sum_{i=1}^{n} \rho_{\tau}( \mathbf{x}_i\trans \hat{\boldsymbol{\beta}}^p - \mathbf{x}_i\trans \boldsymbol{\beta}) , \label{objective_function}
    \end{align}
where $\zeta \in [0,1]$ is a tuning parameter that balances the weight between the quantile check loss of the observed data and that of the prior knowledge. When $\zeta = 1$, the KIQR method relies solely on the prior knowledge; when $\zeta = 0$, our method reduces to a traditional penalized quantile regression model. Note that here $\boldsymbol{\beta}$, $\hat{\boldsymbol{\beta}}^p$, $\rho(\cdot)$, and $\mathbf{Q}(\cdot)$ vary across different values of $\tau$. Again, we omit the subscript $\tau$ for notational simplicity.

The KIQR objective function contains three components: a quantile check loss function for the observed data, a penalty term for variable selection, and a quantile check loss function for incorporating prior knowledge. In addition to the tuning parameter $\lambda$ for the penalty term, a tuning parameter $\zeta$ is introduced to further balance the trade-off between the information derived from prior knowledge and the observed data. Conceptually, the third term resembles knowledge-guided learning in modern AI, in that it uses prior information to anchor the training model and discourage excessive departures from established evidence.
A desirable feature of our approach is the flexibility to accommodate a variety of knowledge formats, such as previously identified important variables, established models, or predicted coefficients from prior studies. Compared to traditional meta-analyses, KIQR is distinguished by its adaptability and flexibility. This flexibility is particularly critical when it becomes necessary to integrate knowledge from domain experts, which is often difficult to include in traditional meta-analytical frameworks. Beyond its inherent flexibility, by proper selection of tuning parameters, KIQR demonstrates robustness to potentially misspecified information from previous studies. 

\subsection{Approximation}     
In the KIQR method, the selection and estimation of important variables via minimizing the objective function \eqref{objective_function} present significant challenges, including the non-differentiable nature of the quantile check loss function and the nonconvex nature of the SCAD penalty function. We tackle these challenges by adopting the Huber loss approximation (HLA, \citealt{Huber1973}) as proposed in \cite{Yi2017} and \cite{Sherwood2022} for the quantile check loss function and the local linear approximation (LLA) as in \cite{Zou2008} for the SCAD penalty function. Although not required to establish theoretical properties, adopting these approximations could substantially improve computational efficiency.

The check loss function $ \rho_{\tau}(u) = u \{ \tau - \mathbf{I} (u < 0) \} $ is equivalent to $\rho_\tau(u)=\frac{1}{2}[|u|+(2 \tau-1) u]$, which is non-differentiable at $u = 0$. We adopt the HLA, which first approximates the non-differentiable absolute value $|u|$ by: 		
    \begin{align}
	g_\gamma(u)= 
	\begin{cases}
	\frac{u^2}{2 \gamma}, & \text { if }|u| \leq \gamma \\ |u|-\frac{\gamma}{2}, & \text { if }|u|>\gamma .
	\end{cases}
    \end{align}
The HLA quantile loss is then defined as:  
    \begin{align}
	h_{\gamma,\tau}(u)=\frac{1}{2}[g_\gamma(u)+(2 \tau-1) u]. 			
    \end{align}	
When $\gamma$ is sufficiently small, $h_{\gamma,\tau}(u) \approx \rho_{\tau}(u)$. In our application, we adopt $\gamma = 0.01$ as suggested in \cite{Sherwood2022}.
	
To tackle the nonconvex nature of the SCAD penalty function, we adopt the LLA method. For any $j$, assuming $\beta_j$ is close to an initial estimate $\hat\beta^{(0)}_{j}$, we can use $\boldsymbol{\phi}^*\left(\beta_j \mid \hat\beta^{(0)}_{j} \right)$ to approximate $p_\lambda\left(|\beta_j|\right)$ by		
    \begin{align}
    \boldsymbol{\phi}^*\left(\beta_j \mid \hat\beta^{(0)}_{j}\right)=p_\lambda\left(|\hat\beta^{(0)}_{j}|\right)+p_\lambda^{\prime}\left(|\hat\beta^{(0)}_{j}|\right)\left(|\beta_j| - |\hat\beta^{(0)}_j|\right).
    \end{align}

Incorporating both the HLA and the LLA, we have the approximated objective function for $\boldsymbol{\beta}$ near $\boldsymbol{\hat\beta}^{(0)}$ as:				
    \begin{align}
	\begin{split}
        \mathbf{H}_{\lambda, \zeta,\gamma}(\boldsymbol{\beta}; \mathbf{X}, \mathbf{y}, \hat{\boldsymbol{\beta}}^p, {\hat{\boldsymbol{\beta}}}^{(0)}) 
	& \approx   \frac{1-\zeta}{n} \sum_{i=1}^{n} h_{\gamma,\tau}(y_i - \mathbf{x}_i \trans \boldsymbol{\beta}) 
			+ (1 - \zeta) \sum_{j=1}^{d} \phi^*\left(\beta_j \mid \hat\beta_j^{(0)}\right)\\
			&  + \frac{\zeta}{n} \sum_{i=1}^{n} h_{\gamma,\tau}(\mathbf{x}_i\trans \hat{\boldsymbol{\beta}}^p - \mathbf{x}_i\trans \boldsymbol{\beta}). \label{Huber_LLA_objective_function}
        \end{split}
    \end{align}
Let  
$h_{\gamma} (\boldsymbol{\beta}) =  \frac{1}{n} \sum_{i=1}^{n} h_{\gamma,\tau}(y_i - \mathbf{x}_i^{\trans} \boldsymbol{\beta})$
and 
$h^{p}_{\gamma}(\boldsymbol{\beta}) =  \frac{1}{n} \sum_{i=1}^{n} h_{\gamma,\tau}( \mathbf{x}_i\trans \hat{\boldsymbol{\beta}}^p - \mathbf{x}_i\trans \boldsymbol{\beta})$.
Again, we omit the subscript $\tau$ and employ a slight abuse of notation for $h$ for notational simplicity. Therefore, the approximated objective function of KIQR can be equivalently written as: 
    \begin{align}
        \label{eqn:objective}
        \mathbf{H}_{\lambda, \zeta, \gamma}(\boldsymbol{\beta}; \mathbf{X}, \mathbf{y},\hat{\boldsymbol{\beta}}^p, {\hat{\boldsymbol{\beta}}}^{(0)})  
        &\approx (1-\zeta)h_{\gamma}(\boldsymbol{\beta}) + \zeta h_{\gamma}^{p}(\boldsymbol{\beta}) + (1 - \zeta) \sum_{j=1}^{d} \phi^*\left(\beta_j \mid \hat\beta_j^{(0)}\right). 
    \end{align}
By minimizing $\mathbf{H}_{\lambda, \zeta, \gamma}$ in Equation~(\ref{eqn:objective}), we obtain the KIQR estimator as: 
    \be
	\wh\vbeta^{KIQR}=\argmin_{\vbeta}\left( (1-\zeta)h_\gamma(\vbeta)+\zeta h_\gamma^p(\vbeta)+ (1 - \zeta)\sum_{j=1}^{d}p'_{\lambda}(| \hat{\beta}^{(0)}_j |)|\beta_j| \right).
	\label{eq:KIQR_estimator}
	\ee
    In our implementation, we employ the traditional penalized quantile regression with the LASSO ($L_1$) penalty for the initial estimate.

\subsection{Computation}
We adopt the cyclic coordinate descent method to minimize the approximated objective function defined in Equation \eqref{Huber_LLA_objective_function}. We initialize the algorithm using the estimate from the traditional penalized quantile regression model. We use ${\hat{\boldsymbol{\beta}}}^{(k)}$ to denote the estimated $\boldsymbol{\beta}$ at the $k$-th iteration. Let us denote $\nabla \mathbf{H}_{\lambda, \zeta, \gamma} (\hat\beta^{(k)}_{j})$ as the first derivative of $\mathbf{H}_{\lambda, \zeta, \gamma} $ with respect to $\hat\beta^{(k)}_{j}$. Therefore, conditional on the $k$-th iteration, for the $j$-th coefficient, we can iteratively update $\hat{\boldsymbol{\beta}}$ as follows:
    \begin{align}
    \label{CD_objective}  
        \boldsymbol{\hat{\beta}}_{j}^{(k+1)} &=
        \underset{\boldsymbol{\beta}_j}{\arg \min } \left\{ \mathbf{H}_{\lambda, \zeta, \gamma}\left(\boldsymbol{\beta}; \mathbf{X}, \mathbf{y}, \hat{\boldsymbol{\beta}}^p, \hat{\boldsymbol{\beta}}^{(k)}\right)  \right\} \nonumber \\  
        \mathbf{H}_{\lambda, \zeta, \gamma}\left(\hat\beta^{(k+1)}_j; \hat\beta^{(k)}_j\right) &\approx \mathbf{H}_{\lambda, \zeta, \gamma}(\hat\beta^{(k)}_j)+\left(\hat\beta^{(k+1)}_j-\hat\beta^{(k)}_j\right) \nabla \mathbf{H}_{\lambda, \zeta, \gamma}(\hat\beta^{(k)}_j)+\frac{D_j}{2}\left(\hat\beta^{(k+1)}_j-\hat\beta^{(k)}_j\right)^2,
    \end{align}
where $D_j = \frac{2}{n \gamma} [\mathbf{X}\trans \mathbf{X}]_{(j,j)} \nonumber$ for ${\hat{\boldsymbol{\beta}}}^{(k+1)}$ close to ${\hat{\boldsymbol{\beta}}}^{(k)}$.
Let $\nabla h_{\gamma} (\hat\beta_j^{(k)})$ denote the first derivative of $h_\gamma(\vbeta) $ with respect to $\hat\beta^{(k)}_{j}$, and $\nabla h^{p}_{\gamma} (\hat\beta^{(k)}_{j})$ denote the first derivative of $h_\gamma^p(\vbeta)$ with respect to $\hat\beta^{(k)}_{j}$.  Taking the first derivative of \eqref{CD_objective} and setting it to zero, we have the estimated $\hat{\beta}_j^{(k+1)}$ as:
	\begin{align}
		\hat{\beta}_j^{(k+1)} &= \hat\beta_j^{(k)}- \frac{1}{D_j}\left[  (1-\zeta) \nabla h_{\gamma} (\hat\beta_j^{(k)}) + \zeta \nabla h^{p}_{\gamma}(\hat\beta_j^{(k)}) + (1 - \zeta)p^{\prime}_{\lambda}(|\hat\beta_j^{(k)}|) \text{sign}(\hat\beta_j^{(k)})   \right],
	\end{align}
	for each coefficient $\beta_j$ of $\boldsymbol{\beta}$, where $j = 1, \dots, d$. We then repeat the fitting process iteratively until the convergence of $\boldsymbol{\beta}$. The pseudocode of this algorithm is shown in Algorithm \ref{alg:KIQR}.

The tuning parameters $\lambda$ and $\zeta$ can be selected by cross-validation. For each $\zeta$ on the grid between 0 and 1, an optimal $\lambda$ is selected using the minimum or 1-SE rule in cross-validation, then among all ($\zeta, \lambda$) combinations, we select the one with the lowest cross-validation score in the check loss. 

Alternatively, motivated by \cite{sherwoodlan15} and \cite{Lee2014}, we can also choose the tuning parameters $\lambda$ and $\zeta$ through the following quantile high dimensional BIC criterion:
    \begin{align*}
        \text{QBIC}(\lambda, \zeta) = & \text{log}\left( \sum_{i = 1}^{n} \rho_{\tau} (Y_i - \mathbf{x}_i^\prime \boldsymbol{\hat{\beta}}_{\lambda, \zeta})  \right) + \nu_{\lambda, \zeta} \frac{\text{log}(d) \text{log}(\text{log}(n))}{2n},
    \end{align*}
where $\nu_{\lambda, \zeta}$ is the number of non-zero coefficients in the fitted model. In our real data analysis and simulations, we employ the QBIC for its computational efficiency compared to cross-validation.

\begin{algorithm}[h]
  \caption{KIQR Estimation and Variable Selection}\label{alg:KIQR}
  \begin{algorithmic}[1]
    \Require The original data set \( (\mathbf{X}, \mathbf{y}) \) and the prior knowledge important variable set \( \mathbf{S_p} \).
    
    \State Obtain the estimator fully trusting prior knowledge \( \hat{\boldsymbol{\beta}}^p \) via penalized quantile regression with no penalty on the variables in $\mathbf{S_p}$, by solving:
    $$
    \hat{\boldsymbol{\beta}}^p = \underset{\boldsymbol{\beta}}{\argmin}
    \left\{ \frac{1}{n} \sum_{i=1}^{n} \rho_{\tau}(y_i - \mathbf{x}_i^\top \boldsymbol{\beta}) + \sum_{j \not\in \mathbf{S}_p} p_{\lambda}(|\beta_j|) \right\}.
    $$
    
    \State Obtain an initial estimator \( \boldsymbol{\hat{\beta}}^{(0)} \) via traditional penalized quantile regression.
    
    \State Apply the cyclic coordinate descent algorithm:
    \State Initialize $k \gets 0$ and set \( \boldsymbol{\hat{\beta}} \gets \boldsymbol{\hat{\beta}}^{(0)} \).
    \Repeat
      \For{$j \gets 1$ to $d$}
        \State Update \( \hat{\beta}_j \):
        $$
        \hat{\beta}_j \gets \hat\beta_j - \frac{1}{D_j}\left[ (1-\zeta) \nabla h_{\gamma}(\hat\beta_j) + \zeta \nabla h^{p}_{\gamma}(\hat\beta_j) + (1-\zeta)p^{\prime}_{\lambda}(|\hat\beta_j|) \text{sign}(\hat\beta_j) \right].
        $$
      \EndFor
    \Until{convergence or reaching the maximum iteration.}

    \State \textbf{Output}: \( \boldsymbol{\hat{\beta}}^{KIQR} \gets \boldsymbol{\hat{\beta}} \).    
  \end{algorithmic}
\end{algorithm}

\section{Obesity Analysis}
\label{sec:realdata}
\subsection{Real Data Analysis}

The central aim of our study is to identify important genetic risk factors contributing to obesity while incorporating prior information from previous studies. Using an ultra-high-dimensional genetic dataset from the Framingham Heart Study data, we apply our proposed KIQR framework, specifically focusing on the high quantiles of BMI. 
	
The FHS dataset we analyzed includes 1,964 participants and 500,568 SNPs, along with covariates such as age and sex, and the primary phenotype variable of interest, BMI. The clinical measurements were collected from 1987 to 1991, while the genome-wide SNP genotyping was conducted starting in 2007 through NHLBI Framingham SNP Health Association Resource (SHARe). The median age of the participants in this study is 49 years. Although the FHS cohort is not intended to be nationally representative, its BMI distribution is broadly comparable to that of the U.S. adult population during the same time period. In particular, the median BMI in our data is 25.9, which is very close to the contemporaneous U.S. adult benchmarks of 25.5 from NHANES III \citep{Kuczmarski1997}, placing it in the adult overweight range (BMI $\ge$ 25).

Of particular clinical concern are the higher BMI quantiles: the 80th percentile of our data is 29.89, close to the clinical obesity threshold of 30. Given that abnormally high BMI, or obesity, is of primary clinical interest, our study specifically targets the upper quantiles of BMI.

In the preprocessing phase of our genetic data, we follow standard quality control guidelines \citep{Marees2018} and exclude SNPs with a minor allele frequency (MAF) below 0.1 or a Hardy-Weinberg equilibrium (HWE) test $p$-value below 0.001. We then conduct a univariate GWAS on BMI, adjusting for sex, age, and the first five genetic principal components to account for population structure \citep{Price2006, Hoffmann2018}. To obtain a computationally feasible yet still ultra-high-dimensional subset for the subsequent KIQR analysis, we retain the top 4,000 SNPs ranked by their marginal univariate $p$-values.

To incorporate valuable prior knowledge from established large-scale studies, we utilize the meta-analysis of BMI-related traits from the UK Biobank and the GIANT consortium \citep{Yengo2018}. After aligning these SNPs with our FHS data, we retain the 103 overlapping SNPs within our FHS dataset that reached genome-wide significance in both sources as our prior set $\mathbf{S}_p$. This construction allows the KIQR model to flexibly integrate external knowledge with the observed FHS dataset while remaining robust to potential misspecifications.

We first focus our analysis on the 80th conditional quantile of BMI to identify the genetic risk factors for obesity through the KIQR approach, which enables us to target risk factors that are highly relevant to individuals at an elevated obesity risk. In addition, we report the results for the conditional median (the 50th percentile) via KIQR. As illustrated in Figure \ref{fig:GWAS}, in contrast to traditional GWAS methods, where no SNPs pass the $p$-value threshold for genome-wide significance ($p$-value = $5\times10^{-8}$), KIQR successfully identifies multiple SNPs at different quantile levels.

\begin{table}[h]
		\centering
		\caption{Obesity Analysis with KIQR. This table shows the selected loci identified at both conditional quantiles ($\tau = 0.8$ and $\tau = 0.5$, Matched), and the selected loci identified exclusively at $\tau = 0.8$ only or $\tau = 0.5$ only (Unmatched). It details the SNPs, their associated genes, the traits they affect, and references confirming these traits.}
		\label{tab:overlap_0.8_0.5}
		\resizebox{\textwidth}{!}{
            \begin{tabular}{llllllll}
				\toprule
				\multicolumn{4}{c}{$\tau = 0.8$} & \multicolumn{4}{c}{$\tau = 0.5$} \\
				\cmidrule(r){1-4} \cmidrule(l){5-8}
				SNP & Gene & Trait & Selected Reference & SNP  & Gene & Trait & Selected Reference \\
				\midrule
				\multicolumn{8}{c}{Matched} \\
				\midrule
                {\bf rs3798696} & {\bf TFAP2A} &  &  & {\bf rs3798696} & {\bf TFAP2A} &  &  \\
				rs417873 &  SLC8A1  &   Body Mass Index  & \cite{Huang2022}    & rs417873  & SLC8A1  &  Body Mass Index &  \cite{Huang2022} \\ 
                rs2838694 &  SUMO3 & Body Height & \cite{Yengo2022} & rs2838694  & SUMO3 &  Body Height & \cite{Yengo2022} \\
                rs10900135 &  TMEM72-AS1 & Body Mass Index & \cite{Anderson2015} & rs10900135  & TMEM72-AS1 &  Body Mass Index & \cite{Anderson2015} \\ 
                rs9366301 &  OFCC1 & Body Mass Index & \cite{Huang2022} & rs9366301  & OFCC1 &  Body Mass Index & \cite{Huang2022} \\ 
                rs2456899 &  RGS5 & Body Height & \cite{Tachmazidou2017} & rs2456899  & RGS5 &  Body Height & \cite{Tachmazidou2017} \\ 
                rs11174225 &  TAFA2 & Body Mass Index & \cite{Zhu2020UKBiobank} & rs11174225  & TAFA2 &  Body Mass Index & \cite{Zhu2020UKBiobank} \\
                rs10519076 &  RORA & Body Mass Index & \cite{Huang2022} & rs10519076  & RORA &  Body Mass Index & \cite{Huang2022} \\
                rs4910316 &  GALNT18 &   &   & rs4910316  & GALNT18 &    &   \\
                rs3087253 &  CCR5AS & Body Height & \cite{Yengo2022} & rs3087253  & CCR5AS &  Body Height & \cite{Yengo2022} \\
				\midrule
				\multicolumn{8}{c}{Unmatched} \\
				\midrule
				rs9295657  & CARMIL1 & Body Mass Index &\cite{Hawkes2023} & rs7076242 & ASAH2  & Body Height &  \cite{Yengo2022}  \\
				rs12696582  & LPP & Body Mass Index & \cite{Huang2022} & rs6463489  & FBXL18 & Body Mass Index & \cite{Huang2022}  \\
				rs178260  & \textbf{AIFM3} & \textbf{Early on-set extreme obesity} & \cite{Loid2022} & rs11686962  & LTBP1 & Body Mass Index & \cite{Christakoudi2021}  \\
				rs959247  & C4orf54 & Body Height & \cite{Yengo2022} & rs551167  & DDX25 &   &    \\
				rs17579544  & GDA & Body Height & \cite{Yengo2022} & rs11536450 & NOD1  &   &   \\
				rs4440231  & PDS5A & Body Height & \cite{Schoeler2023} & rs868554  & GPR139 & Body Mass Index & \cite{Huang2022} \\
				\textbf{rs7070523}  & \textbf{ITIH5} &   &    & rs11611246  & WNK1 & Body Mass Index & \cite{Huang2022}   \\
				rs7705511  & CTNND2 & Body Mass Index & \cite{Jee2025} & rs1006107  & HECW1 & Body height & \cite{Yengo2022}  \\
				rs7020239  & PTPRD & Body Mass Index & \cite{Huang2022} & rs2549161  & CDH13 & Body Mass Index & \cite{Huang2022}  \\
                rs6477694  & EPB41L4B & Obesity & \cite{Mokry2016} & rs11652148  & HOXB3 & Obesity & \cite{Verma2024}  \\
                
				\bottomrule
			\end{tabular}
            }
    \end{table}

Table \ref{tab:overlap_0.8_0.5} shows the top candidate loci identified at both conditional quantiles ($\tau = 0.8$ and $\tau = 0.5$), as well as the non-overlapping (unmatched) top candidate loci. Importantly, the KIQR approach uncovers novel associations that have not been previously reported in such large-scale GWAS studies.
For instance, the KIQR method identifies rs3798696 in \textit{TFAP2A} in both $\tau = 0.8$ and $\tau = 0.5$ (indicated by the black triangle in Figure~\ref{fig:GWAS}) with the largest effect size, although this SNP has never been reported by large-scale GWAS as being directly associated with BMI or body-shape-related traits. However, this finding is corroborated by independent epigenetic evidence. For example, \cite{Noor2019} reveal through motif-finding analysis that \textit{TFAP2A} binding motifs are enriched in the sequences surrounding the CpG sites associated with both paternal and maternal BMI. Furthermore, \cite{Scott2018} and \cite{Baca2022} demonstrate that \textit{TFAP2A} is a key regulator of lipid droplet accumulation, offering a plausible biological pathway for its influence on BMI. 

At only $\tau = 0.8$, KIQR identifies rs7070523 in \textit{ITIH5} (marked by a red triangle in Figure~\ref{fig:GWAS}), which has not been reported by large-scale GWAS as being directly associated with BMI or body shape-related traits. This finding is consistent with the observed high expression of \textit{ITIH5} in adipose tissue reported by \cite{Anveden2012}. Furthermore, \cite{Dahlman2012} support this finding by highlighting the role of \textit{ITIH5} in the extracellular matrix, which is critical for obesity-associated adipose inflammation and insulin resistance.

Notably, again only at $\tau = 0.8$, KIQR identifies rs178260 in \textit{AIFM3} (marked by a red triangle in Figure~\ref{fig:GWAS}), a locus not previously reported in large-scale GWAS studies of BMI or related body-shape traits. Although absent in general population studies, \textit{AIFM3} was highlighted by \cite{Loid2022} as a candidate gene for early-onset extreme obesity following the discovery of rare missense variants in affected individuals. This specifically supports our finding and suggests that \textit{AIFM3} possesses a heterogeneous effect, acting primarily on individuals at the upper tail of the BMI distribution.\footnote{
KIQR can be similarly applied to different quantiles of interest. For example, when applying to $\tau=0.9$, we find that rs6477694 in gene EPB41L4B, confirmed directly associated with obesity through Mendelian Randomization Study \citep{Mokry2016}, is selected for both $\tau = 0.8, 0.9$ but not $\tau=0.5$. On the other hand, only at $\tau = 0.9,$ rs7334078 in gene STK24, confirmed association with BMI \citep{Huang2022}, is selected but not at $\tau=0.5, 0.8$. This may warrant future investigation.}

Finally, the KIQR method successfully recovers the top signals detected by traditional GWAS despite the limited sample size (e.g., rs2456899 in \textit{RGS5} at Chromosome 1 and rs10519076 in \textit{RORA} at Chromosome 15, as shown in Figure \ref{fig:GWAS}), thereby confirming the validity and reliability of our approach. Many of the top loci can also be verified by large-scale GWAS studies.\footnote{For example, rs417873 in \textit{SLC8A1}, rs9295657 in \textit{CARMIL1}, and rs6463489 in \textit{FBXL18} can be verified by previous GWAS studies to be associated with BMI, as detailed in Table \ref{tab:overlap_0.8_0.5}.}

\subsection{Mimic Data Simulations} \label{sec:mimic}

We further conduct a mimic genetic data simulation study to evaluate the performance of our proposed KIQR method. We randomly select $n=250$ subjects and $d=1,500$ SNPs from the FHS data. The resulting design matrix also includes age and sex as covariates. The response variable is then generated according to the model $y_i = \sum_{j = 1}^{d}x_{ij} {\beta_j}+ \epsilon_i$, where the first 20 coefficients (corresponding to age, sex, and 18 SNPs) are set to 1, and all others are 0. The error terms $\epsilon_i$ are generated from an i.i.d. normal distribution with a standard deviation of 0.6.

We compare the performance of KIQR against three benchmark models: (1) a GWAS-style marginal quantile regression ({GWAS-QR})\footnote{Traditional GWAS tests SNP effects using the $p$-values from marginal univariate mean regression, while GWAS-QR extends this framework using marginal univariate quantile regression, allowing for the assessment of genetic effects across the distribution of BMI.} similar to \cite{Wang2024}, where each SNP is tested individually with age and sex as covariates; (2) a standard penalized quantile regression without any prior knowledge ({Trad-QR}); and (3) a penalized quantile regression that fully trusts the prior knowledge by not penalizing the variables in the prior set ({Prior-QR}). For the GWAS-QR method, we adopt the widely used genome-wide significance $p$-value threshold of $5 \times 10^{-8}$ as the cutoff to select the important SNPs. To assess KIQR's ability to leverage useful information while remaining robust to flawed data, we test four scenarios for the prior knowledge set: \\
    \text{S1}: perfect prior knowledge, containing all 20 true predictors;\\
    \text{S2}: mostly correct prior, with 10 true predictors and 2 false predictors;\\
    \text{S3}: mixed-quality prior, with 5 true and 15 false predictors;\\
    \text{S4}: completely incorrect prior, with 20 false predictors.
    
The simulation is replicated 200 times, and we present the results for two quantile levels, $\tau = 0.5$ and $\tau = 0.8$. Performance is evaluated based on True Positives (TP), False Positives (FP), the F1 score, and the Mean Squared Error (MSE). The F1 score, a measure of variable selection performance, is defined as $2\text{TP}/(2\text{TP}+\text{FP}+\text{FN})$, where FN is the number of false negatives. The MSE is the mean squared difference between the estimated and true coefficient vectors.

Figure~\ref{fig:mimic} shows the notably superior performance of the KIQR method over GWAS-QR and traditional penalized quantile regression (Trad-QR). Overall, KIQR substantially outperforms the baseline GWAS-QR approach across all scenarios. With informative and correct knowledge, as shown in scenario S1, KIQR and Prior-QR demonstrate superior performance over Trad-QR at both quantile levels, achieving a higher F1 score compared to Trad-QR. In scenarios with a mix of correct and incorrect prior variables, such as S2 and S3, KIQR still outperforms Trad-QR. Even in the extreme scenario S4, where all 20 priors are incorrect and Prior-QR yields poor variable selection results, KIQR maintains robust performance against misleading prior information. Table \ref{tab:mimic_results_new} presents detailed results in terms of F1 score, TP, FP, and MSE. Again, at both the upper quantile ($\tau=0.8$) and the median ($\tau=0.5$), the KIQR model shows superior performance and robustness to misspecified prior information.

\begin{figure}[h!]
    \centering
    \caption{F1 scores in the mimic data simulation for $\tau$ = 0.5 and 0.8 with a normally distributed error term, using a GWAS-style marginal quantile regression ({GWAS-QR}), a penalized quantile regression without prior knowledge (Trad-QR),  a penalized quantile regression with fully trusting the prior knowledge in 4 different scenarios (Prior-QR), and KIQR in 4 different scenarios.}
    \includegraphics[width=0.95\linewidth]{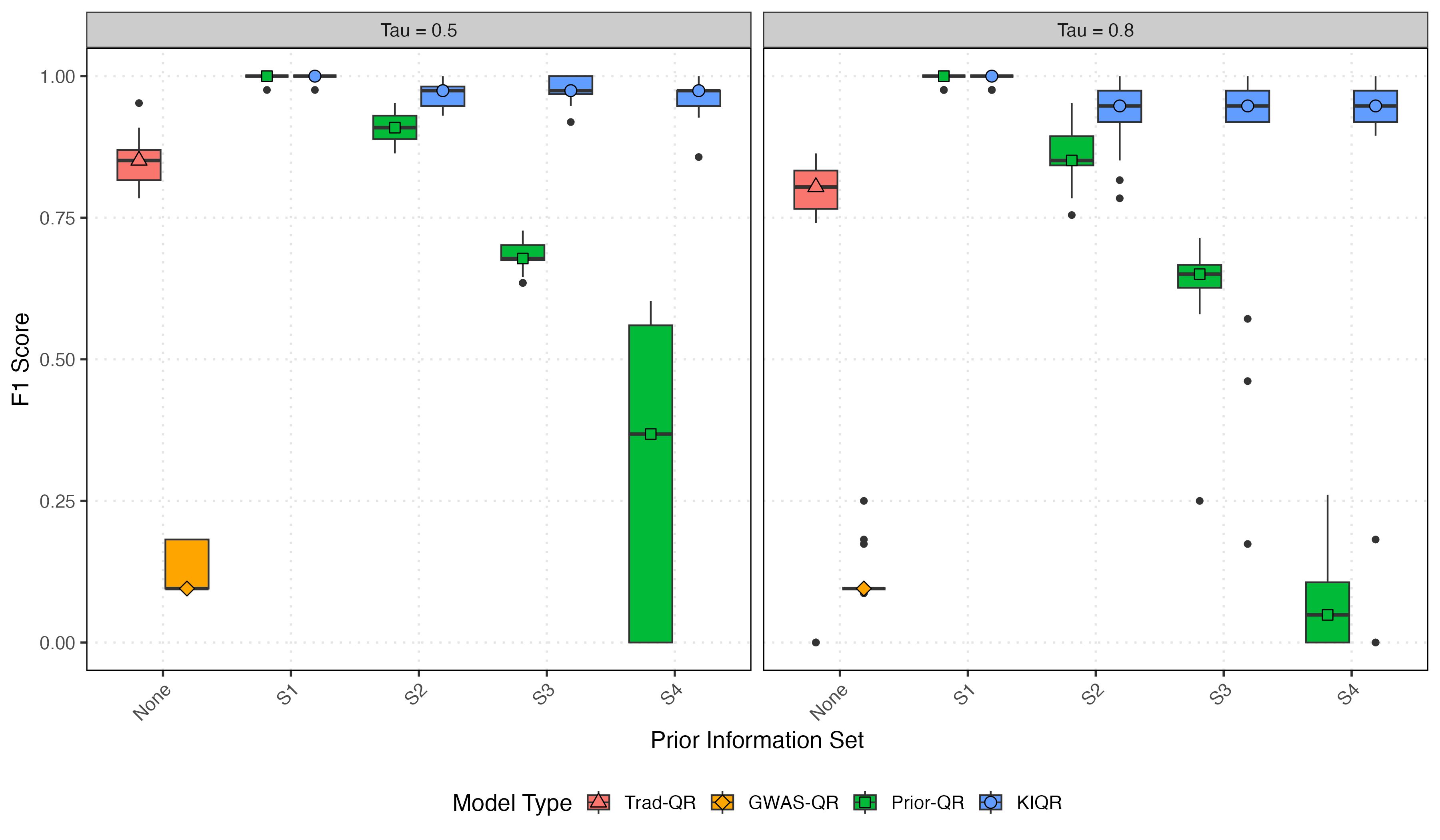}
    \label{fig:mimic}
\end{figure}

\begin{table}[h!]
\centering
\caption{Performance comparison in the mimic data simulation. The table reports the average F1 score, True Positives (TP), False Positives (FP), and Mean Squared Error (MSE) across 200 replications.}
\label{tab:mimic_results_new}
\begin{tabular}{@{}l l rrrr rrrr@{}}
\toprule
& & \multicolumn{4}{c}{$\boldsymbol{\tau = 0.8}$} & \multicolumn{4}{c}{$\boldsymbol{\tau = 0.5}$} \\
\cmidrule(lr){3-6} \cmidrule(lr){7-10}
\textbf{Prior} & \textbf{Model} & \textbf{F1} & \textbf{TP} & \textbf{FP} & \textbf{MSE} & \textbf{F1} & \textbf{TP} & \textbf{FP} & \textbf{MSE} \\
\midrule
None & GWAS-QR & 0.195 & 2.20 & 0.40 & 0.0127 & 0.209 & 2.35 & 0.10 & 0.0122 \\
& Trad-QR & 0.754 & 16.90 & 7.95 & 0.0032 & 0.850 & 19.90 & 6.90 & 0.0012 \\
\midrule
S1 & KIQR & 0.996 & 20.00 & 0.15 & 0.0002 & 0.999 & 20.00 & 0.05 & 0.0001 \\
\makecell[l]{} & Prior-QR & 0.996 & 20.00 & 0.15 & 0.0002 & 0.999 & 20.00 & 0.05 & 0.0001 \\
\midrule
S2 & KIQR & 0.934 & 18.70 & 1.35 & 0.0012 & 0.971 & 19.10 & 0.25 & 0.0009 \\
\makecell[l]{} & Prior-QR & 0.859 & 20.00 & 6.55 & 0.0007 & 0.909 & 19.90 & 3.90 & 0.0007 \\
\midrule
S3 & KIQR & 0.903 & 16.50 & 0.05 & 0.0027 & 0.977 & 19.15 & 0.05 & 0.0009 \\
\makecell[l]{} & Prior-QR & 0.633 & 18.80 & 20.60 & 0.0027 & 0.679 & 20.00 & 18.90 & 0.0012 \\
\midrule
S4 & KIQR & 0.881 & 15.80 & 0.05 & 0.0030 & 0.964 & 18.80 & 0.20 & 0.0011 \\
\makecell[l]{} & Prior-QR & 0.070 & 1.45 & 20.00 & 0.0148 & 0.337 & 8.50 & 22.00 & 0.0104 \\
\bottomrule
\end{tabular}
\end{table}

\section{Numerical Simulations}	\label{sec:sim}
\subsection{Example 1}
To further assess the performance of the KIQR method in variable selection and estimation across different quantiles for ultra-high-dimensional data, we conduct the following numerical simulations.\footnote{This work received computational support from the UTSA Arc HPC environment, operated by University Tech Solutions.}
    
In our simulations, we first employ a data generating process for $i = 1, \dots, n$, given by
$y_i = \sum_{j = 1}^{d}x_{ij} {\beta_j}+ \epsilon_i, $
where the design matrix $\mathbf{X}$ follows a multivariate normal distribution with mean $\mathbf{0}$ and covariance matrix $\mathbf{\Sigma}$. Here, $\mathbf{\Sigma} = (\sigma_{ij})$ follows an AR(1) covariance structure where $\sigma_{ij} = \rho^{|i - j|}$ for $i,j = 1, \dots, d$, and we set $\rho = 0.5$. The coefficient vector $ \boldsymbol{\beta} $ is set to $ (1, \dots, 1, 0, \ldots, 0)\trans $, where $\beta_{1}, \dots, \beta_{20} = 1$. The error terms $ \epsilon_i $ are generated from either (a) a standard normal distribution of $N(0,1)$, or (b) a $t$ distribution with degrees of freedom 3. We set the number of covariates at a high dimension of $d = 1,500$ and a sample size of $n = 200$. 

To assess the performance of variable selection, we evaluate the TP, FP, and F1 score as defined in Section \ref{sec:mimic} based on 200 simulation runs. To assess the KIQR model's ability to incorporate valuable prior knowledge as well as its robustness against potential erroneous prior information, we again examine the four scenarios S1, S2, S3, and S4 as defined in Section \ref{sec:mimic}. We apply and report the results of the KIQR method at quantile levels of $\tau = 0.5$ and $0.8$, to investigate the model's performance across different quantiles, especially for the high quantiles.
	
We compare the performance of the KIQR method with traditional penalized quantile regression without prior knowledge (Trad-QR). We also benchmark against a penalized quantile regression estimator that fully trusts the prior knowledge (Prior-QR) based on Equation~(\ref{betap}). 

\begin{figure}[h!]
    \centering
    \caption{F1 scores in simulation Example 1 for $\tau$ = 0.5 and 0.8 with a normally distributed error term, using a penalized quantile regression without prior knowledge (Trad-QR),  a penalized regression that fully trusts the prior knowledge in 4 different scenarios (Prior-QR), and KIQR in 4 different scenarios.}
    \includegraphics[width=1\linewidth]{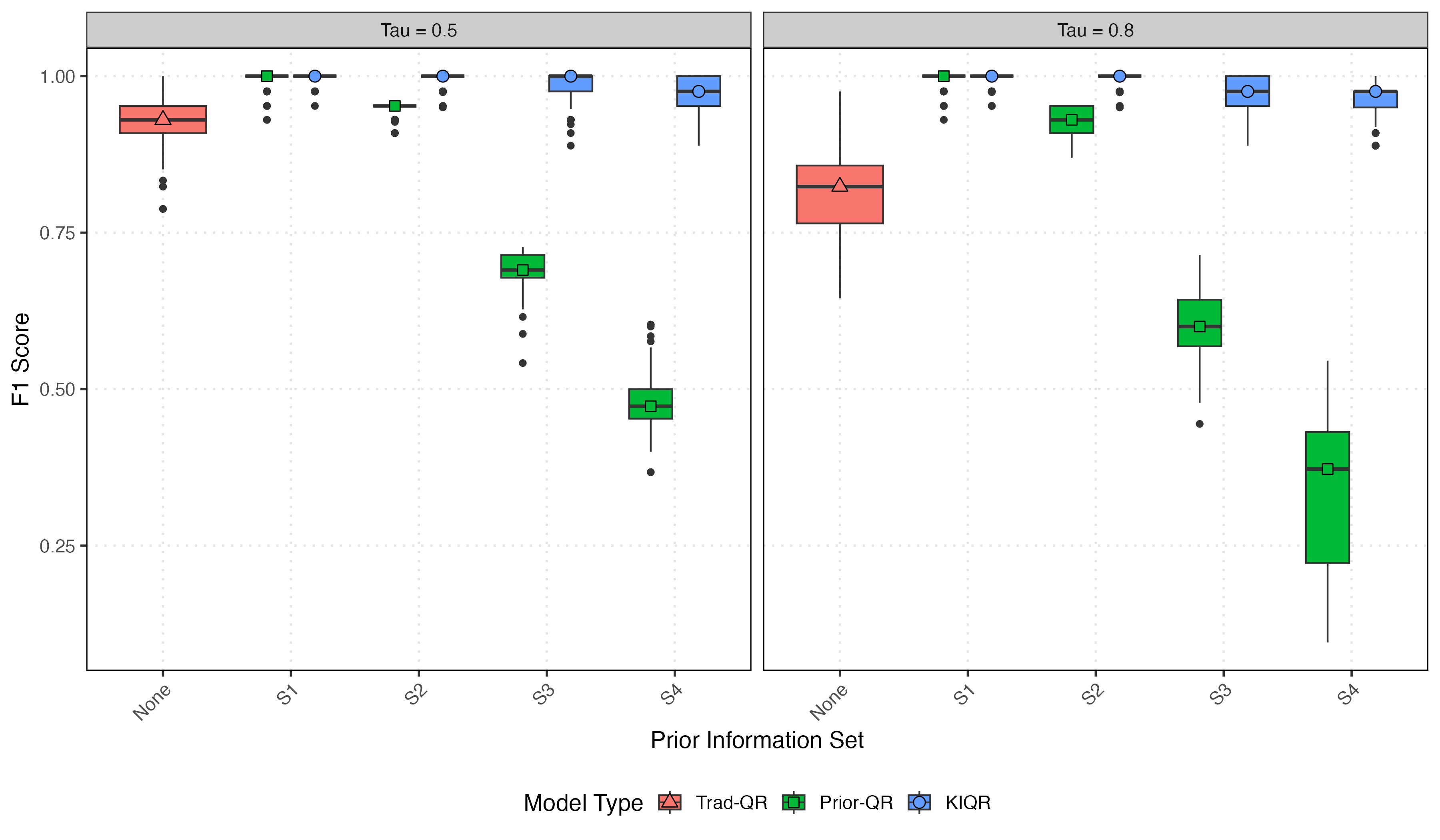}
    \label{fig:f1-score-sim-norm}
\end{figure}
\begin{figure}[h!]
    \centering
    \caption{F1 scores in simulation Example 1 for $\tau$ = 0.5 and 0.8 with a \textit{t-}distributed error term, using a penalized quantile regression without prior knowledge (Trad-QR),  a penalized regression that fully trusts the prior knowledge in 4 different scenarios (Prior-QR), and KIQR in 4 different scenarios.}
    \includegraphics[width=1\linewidth]{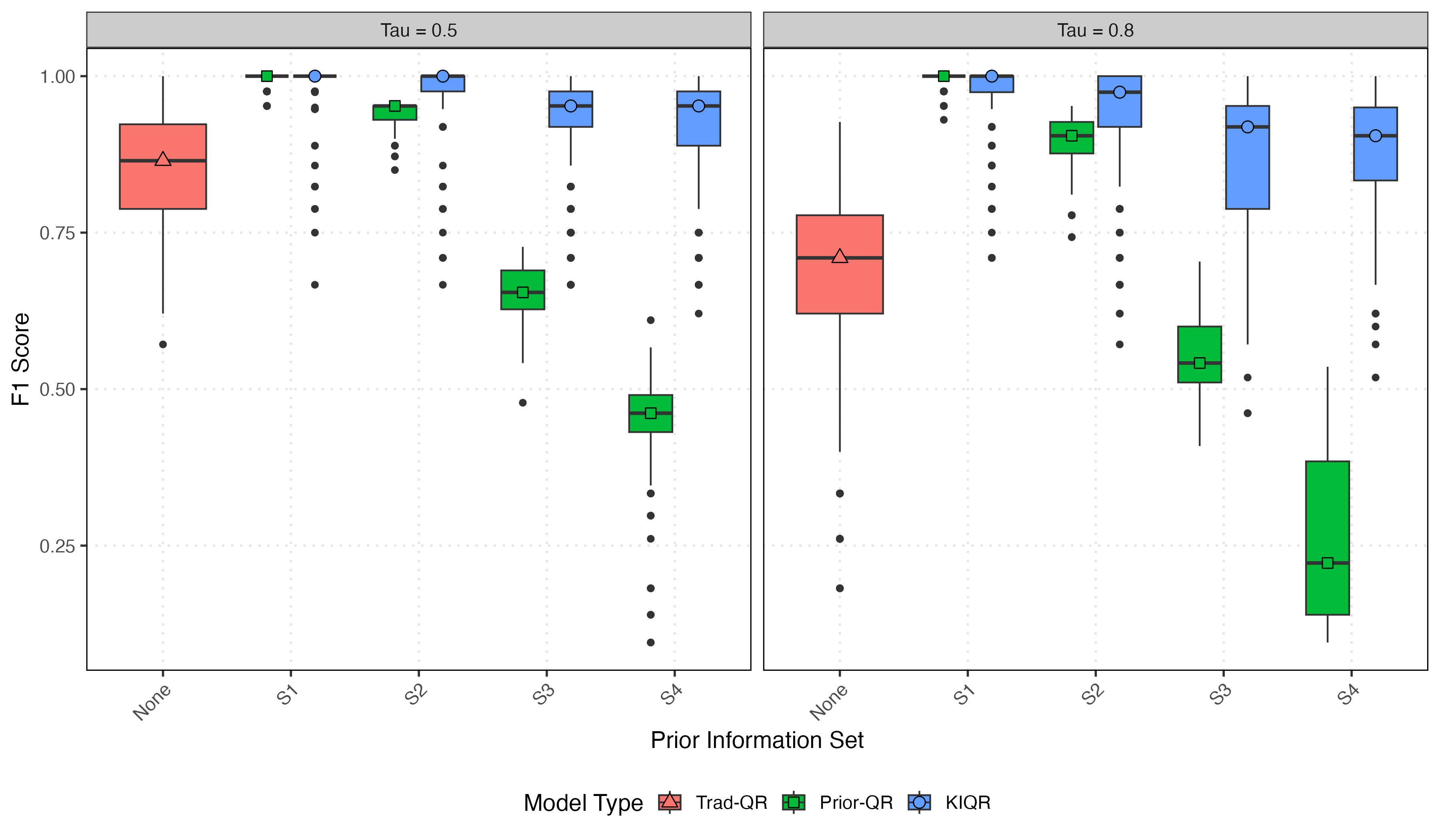}
    \label{fig:f1-score-sim-t3}
\end{figure}

In Example 1, as shown in Figures \ref{fig:f1-score-sim-norm} and \ref{fig:f1-score-sim-t3}, we observe that KIQR outperforms Trad-QR in all four scenarios. With informative and correct knowledge, as shown in scenario S1, KIQR demonstrates superior performance over Trad-QR at both quantile levels, performing comparably to Prior-QR. In scenarios with mixed knowledge of correct and incorrect prior variables, such as in S2 and S3, KIQR still outperforms Trad-QR. In the extreme case of S4, where all prior knowledge is incorrect and Prior-QR fails, KIQR's performance remains superior to that of Trad-QR, indicating KIQR's robustness against misleading prior information.

\subsection{Example 2}
We further conduct simulations in a more challenging setting, based on a model employed by \cite{Wang2012}. We first generate $\tilde{\mathbf{X}}$ from a multivariate normal distribution with mean $\mathbf{0}$ and covariance matrix $\mathbf{\Sigma}_{ij} = (\sigma_{ij})$ where $\sigma_{ij} = $ $\rho^{|i - j|}$ with $\rho = 0.5$. The design matrix $\mathbf{X}$ is then constructed as $X_1 = \Phi(\tilde{X}_1),$ where $\Phi$ is the cumulative distribution function of a standard normal random variable, and $X_j = \tilde{X}_j$ for $j \neq 1$. Subsequently, for $i = 1, \dots, n,$ we generate the response variable from 
	$	y_i =  \sum_{j = 1}^{d}x_{ij} {\beta_j} +  x_{i1} \epsilon_i. $ 
We maintain the high dimensional covariate setting of $d = 1,500$ and a sample size of $n = 200$. The coefficient vector $ \boldsymbol{\beta} $ is defined such that $\beta_{j} = 1$ for $j\in \{10,20, \dots, 50\}$, and $\beta_{j} = 0$ otherwise. In this setting, $X_1$ does not affect the median of $y$ when $\tau = 0.5$, while $X_1$ is an important variable for $\tau \neq 0.5$. The error terms are drawn from a standard normal distribution, $ \epsilon_i \sim N(0, 1)$. The same prior knowledge scenarios (S1, S2, S3, and S4) defined in Example 1 are utilized here.

We present the results in Table \ref{tab:exp2}.
For the interesting upper quantile ($\tau = 0.8$), the superior performance of KIQR is evident. Our proposed KIQR method achieves True Positives (TPs) of 18.95, 18.64, 17.74, and 17.79 across the four different scenarios of prior knowledge, which are higher than the 12.38 achieved by the traditional penalized quantile regression that does not incorporate prior knowledge. 
Regarding False Positives (FPs), KIQR consistently produces fewer FPs compared to the traditional QR method, regardless of the quality of prior knowledge. 
When the prior knowledge is informative (S1 and S2), KIQR yields lower FP counts of 0.09 and 0.14, respectively, compared to 1.17 for the traditional QR. When the prior knowledge is largely misleading (S3 and S4), the FP counts for KIQR (0.81 and 1.21) are comparable to those of the standard QR, again highlighting the robustness of our method against misspecified priors. Furthermore, for $\tau=0.5,$ the traditional QR method does not select the heteroscedastic variable $X_1$ at all. Moreover, at $\tau=0.8,$ Trad-QR barely identifies $X_1$, selecting it in only 2.0\% of replicates. In contrast, when the prior information contains $X_1$ as in scenario $S_1$, KIQR identifies it in a majority of replicates (67.0\%). 

In summary, these results further demonstrate that KIQR delivers superior performance when prior knowledge is informative and maintains robustness against potentially misleading prior information.

\begin{table}[h!]
\centering
\caption{\label{tab:exp2}Simulation results for the high-dimensional heteroscedastic model ($n=200, p=1500$) in Example 2. The model includes a predictor, $X_1$, which is only active for $\tau \neq 0.5$. The table reports the average F1 score (F1), True Positives (TP), False Positives (FP), Mean Squared Error (MSE), and the selection rate for $X_1$ across various prior information sets. F1, TP, FP, and MSE are calculated excluding $X_1$ to evaluate performance on the main effect variables.}
\centering
\begin{tabular}[t]{llrrrrrr}
\toprule
\multicolumn{3}{c}{ } & \multicolumn{4}{c}{Performance (excluding $X_1$)} & \multicolumn{1}{c}{ } \\
\cmidrule(l{3pt}r{3pt}){4-7}
Tau & Prior & Model & F1 & TP & FP & MSE & $X_1$ Sel. (\%)\\
\midrule
0.8 & None & Trad-QR & 0.751 & 12.38 & 1.17 & 0.0070 & 2.0\\
\cmidrule{2-8}
0.8 & S1 & KIQR & 0.996 & 18.95 & 0.09 & 0.0002 & 67.0\\
 & & Prior-QR & 0.994 & 19.00 & 0.25 & 0.0002 & 100.0\\
\cmidrule{2-8}
0.8 & S2 & KIQR & 0.986 & 18.64 & 0.14 & 0.0006 & 42.5\\
 & & Prior-QR & 0.918 & 18.57 & 2.91 & 0.0009 & 100.0\\
\cmidrule{2-8}
0.8 & S3 & KIQR & 0.942 & 17.74 & 0.81 & 0.0015 & 9.5\\
 & & Prior-QR & 0.585 & 14.45 & 15.74 & 0.0066 & 1.0\\
\cmidrule{2-8}
0.8 & S4 & KIQR & 0.935 & 17.79 & 1.21 & 0.0014 & 10.5\\
 & & Prior-QR & 0.320 & 7.76 & 20.55 & 0.0138 & 0.0\\
\midrule
0.5 & None & Trad-QR & 0.932 & 18.56 & 2.29 & 0.0009 & 0.0\\
\cmidrule{2-8}
0.5 & S1 & KIQR & 1.000 & 19.00 & 0.01 & 0.0001 & 2.0\\
 & & Prior-QR & 0.994 & 19.00 & 0.23 & 0.0001 & 100.0\\
\cmidrule{2-8}
0.5 & S2 & KIQR & 0.999 & 19.00 & 0.05 & 0.0001 & 0.5\\
 & & Prior-QR & 0.942 & 18.97 & 2.31 & 0.0002 & 100.0\\
\cmidrule{2-8}
0.5 & S3 & KIQR & 0.993 & 18.98 & 0.24 & 0.0001 & 0.0\\
 & & Prior-QR & 0.687 & 18.43 & 16.16 & 0.0015 & 0.0\\
\cmidrule{2-8}
0.5 & S4 & KIQR & 0.977 & 18.93 & 0.84 & 0.0002 & 0.0\\
 & & Prior-QR & 0.497 & 13.57 & 21.37 & 0.0077 & 0.0\\
\bottomrule
\end{tabular}
\end{table}

\section{Theoretical Properties} \label{sec:theory}

In this section, we establish the oracle properties of the estimator derived from the KIQR method, utilizing the Huber loss approximation (HLA) for the quantile check loss function and the local linear approximation (LLA) for the SCAD penalty. Again, we omit the subscript $\tau$ for different conditional quantiles throughout this section for simplicity. 

We first establish an oracle inequality for the KIQR estimator with a LASSO penalty, defined as:
$$
\wh\vbeta^{KIQR0}=\argmin_{\vbeta}(1-\zeta)h_\gamma(\vbeta)+\zeta h_\gamma^p(\vbeta)+ (1 - \zeta)\lambda\|\vbeta\|_1.
\label{eq:KIQR0_estimator}
$$
Then, starting from the KIQR estimator with the LASSO penalty, we show that the KIQR estimator with the SCAD penalty, $\hat{\boldsymbol{\beta}}^{KIQR}$, defined in Equation~(\ref{eq:KIQR_estimator}), has the appealing oracle property. Note that the values of $\lambda$ in the above two equations may differ. In the following analysis, $C$ denotes a generic positive constant which may assume different values at different places. The following assumptions are imposed.
    \begin{itemize}
		\item[(A1)] $\vx_i$ is sub-Gaussian in the sense that $E\exp\{\vx_i\trans\va\}\le C\exp\{C\|\va\|^2\}$ for any vector $\va$. The matrix $\vSigma=E[\vx_i\vx_i\trans]$ has eigenvalues bounded and bounded away from zero.
		\item[(A2)] The conditional density of $\ep_i=y_i-\vx_i\trans\vbeta_0$, $f_\ep(.|\vx)$, is uniformly bounded by a constant $\bar f$, and its derivative is uniformly bounded by another constant $\bar{f'}$. On any interval $[-C,C]$, $f_\ep(.|\vx)$ is bounded below by a constant $\underline{f}>0$ (where $\underline{f}$ depends on $C$ but this dependence is suppressed in the notation for simplicity).
		\item[(A3)] The true parameter $\vbeta_0$ is sparse, with a support set $S=\{j:\beta_{0j}\neq 0\}$ and $|S|=s_n$. 
    \end{itemize}

    \begin{thm}\label{thm:1}
		Under assumptions (A1)-(A3), and further assuming that $\|  \hat{\vbeta}^p-\vbeta_0\|\le c\gamma $ for a sufficiently small constant $c$, with $\lambda\ge 2(1-\zeta)\|\nabla h_\gamma(\vbeta_0)\|_\infty+2\zeta\|\nabla h_\gamma^p(\vbeta_0)\|_\infty$, and  $\gamma>>\sqrt{s_n\log(d \vee n)/n}$, $\gamma>>\lambda\sqrt{s_n}$, we have with probability at least $1-(d \vee n)^{-C}$ that the estimator of Equation~(\ref{eq:KIQR0_estimator}) satisfies the following oracle inequalities:
		\bse
		\|\wh\vbeta^{KIQR0}-\vbeta_0\|\le C\lambda\sqrt{s_n}\; \mbox{ and } \|\wh\vbeta^{KIQR0}-\vbeta_0\|_1\le C\lambda s_n.
		\ese
    \end{thm}

    \begin{rmk}
		By Theorem \ref{thm:1}, we can take $\lambda=C(1-\zeta)\sqrt{\frac{\log(d\vee n)}{n}}+2\zeta\|\nabla h_\gamma^p(\vbeta_0)\|_\infty$. For the standard LASSO penalized quantile regression without the $h_\gamma^p$ term for prior knowledge, we would take $\lambda=C\sqrt{\frac{\log(d \vee n)}{n}}$ which would also lead to $\|\wh\vbeta^{KIQR0}-\vbeta_0\|\le C\lambda\sqrt{s_n}$ (here $\wh\vbeta^{KIQR0}=\wh\vbeta^{qLASSO}$ is the quantile LASSO estimator without using $h_\gamma^p$, equivalent to setting $\zeta=0$). Thus if $\|\nabla h_\gamma^p(\vbeta_0)\|_\infty$ is of a smaller order than $\sqrt{\frac{\log(d \vee n)}{n}}$, the KIQR estimator uses a smaller $\lambda$ and thus the upper bound is smaller. This observation is similar to that made in \cite{Jiang2016}.
    \end{rmk}

The proof of Theorem \ref{thm:1} utilizes the following lemmas.

    \begin{lem}
    \label{lem:convex}  Let $c>0$ be a sufficiently small constant. Under the same assumptions used in Theorem \ref{thm:1},
		uniformly over the set $\Omega:=\{(\vbeta_1,\vbeta_2): \|\vbeta_1-\vbeta_2\|\le c\gamma, \|\vbeta_2-\vbeta_0\|\le c, \|\vbeta_2-\vbeta_0\|_1\le 3\sqrt{s_n}\|\vbeta_2-\vbeta_0\|\}$, we have with probability at least $1-e^{-t}$,
		\bse
		&&h_\gamma( \vbeta_1)-  h_\gamma( \vbeta_2)- \langle \nabla h_\gamma( \vbeta_2),\vbeta_1-\vbeta_2\rangle\\
		&\ge& \left(\frac{\underline{f}\lambda_{\min}(\vSigma)}{4}    - \frac{C\sqrt{s_n}}{\gamma}\left( \sqrt{\frac{\gamma\log (d)}{n}}+\frac{\log (d)}{n} \right)-C\sqrt{\frac{t}{n\gamma}}-C\frac{t}{n\gamma}  \right) \|\vbeta_1-\vbeta_2\|^2,
		\ese
		where $\lambda_{\min}(\vSigma)$ is the minimum eigenvalue of $\vSigma=E[\vx_i\vx_i\trans]$.
    \end{lem}

    \begin{lem}\label{lem:convex2} 
		Assume $c$ is a sufficiently small positive constant. Under the same assumptions used in Theorem \ref{thm:1}, we have that, with probability at least $1-e^{-t}$, uniformly over the set $\Omega'=\{(\vbeta_1,\vbeta_2):\|\vbeta^p-\vbeta_2\|\le c\gamma, \|\vbeta_1-\vbeta_2\|\le c\gamma,\|\vbeta_1-\vbeta_2\|_1\le 3\sqrt{s_n}\|\vbeta_1-\vbeta_2\| \}$, we have
		\bse
		&&h_\gamma^p( \vbeta_1)- h_\gamma^p(\vbeta_2)-  \langle \nabla h_\gamma^p(\vbeta_2),\vbeta_1-\vbeta_2\rangle \\ &\ge&  \left(\frac{\lambda_{\min}(\vSigma)}{2}-C\left(\frac{1}{\gamma}  \sqrt{\frac{ s_n\log (d)}{n}}   +\sqrt{\frac{t}{n\gamma^2}}+\frac{t}{n\gamma}\right)    \right)\|\vbeta_1-\vbeta_2\|^2.
		\ese
    \end{lem}
	
    \begin{lem}\label{lem:nablah}
		With probability at least $1-e^{-t}$,
		\bse
		\|\nabla h_\gamma(\vbeta_0)\|_\infty\le C\left(\gamma^2+\sqrt{\frac{t+\log (d)}{n}}+\frac{t+\log (d)}{n}\right).
		\ese
    \end{lem}

We now establish the oracle property for the KIQR estimator incorporating the Huber loss approximation for the quantile check loss and the local linear approximation for the SCAD penalty.

Define $\wh\vbeta^{ora}=(\wh\vbeta^{ora}_S,\vnull)=\argmin_{\vbeta:\vbeta_{S^c}=\vnull}(1-\zeta)h_\gamma(\vbeta)+\zeta h_\gamma^p(\vbeta)$. Theorem \ref{thm:2} shows the oracle property that, $\hat{\boldsymbol{\beta}}^{KIQR}$ defined in Equation~(\ref{eq:KIQR_estimator}), using $\wh\vbeta^{KIQR0}$ as the initial estimator, is equal to $\wh\vbeta^{ora}$ with probability approaching one. Here we require the smallest signal to be sufficiently large, a standard condition required for the oracle property to hold in SCAD-penalized models.

    \begin{thm}\label{thm:2}
		Assume $\|\wh\vbeta^{ora}-\vbeta_0\|\le a_n$ and $\|\wh \vbeta^{KIQR0} -\vbeta_0\|\le b_n$ for some positive sequences $a_n,b_n=o(1)$. If $b_n<<\lambda<<\min_{j\in S}|\beta_{0j}|$ and $\lambda>(1-\zeta)\|\nabla h_\gamma(\hat\vbeta^{ora})\|_\infty+\zeta\|\nabla h_\gamma^p(\wh\vbeta^{ora})\|_\infty$, for sufficiently large  $n$, we have that the KIQR estimator defined in Equation~(\ref{eq:KIQR_estimator}) is precisely the oracle estimator
		$$\wh\vbeta^{KIQR}=\wh\vbeta^{ora}.$$
    \end{thm}

    \begin{rmk} 
    In the statement of Theorem \ref{thm:2}, we condition on $\|\wh\vbeta^{ora}-\vbeta_0\|\le a_n$, $\|\wh\vbeta^{KIQR0}-\vbeta_0\|\le b_n$ and the magnitude of $(1-\zeta)\|\nabla h_\gamma(\wh\vbeta^{ora})\|_\infty+\zeta\|\nabla h_\gamma^p(\wh\vbeta^{ora})\|_\infty$; therefore, the statement is entirely deterministic.
		The rate of $\|\wh\vbeta^{ora}-\vbeta_0\|$ inherently depends on that of $\|\wh\vbeta^p-\vbeta_0\|$. If $\wh\vbeta^p$ substantially deviates from the true $\vbeta_0$, we cannot expect $\wh\vbeta^{ora}$ to be a good estimator unless $\zeta=0$. On the other hand, if $\|\wh\vbeta^p-\vbeta_0\|=O_p(\sqrt{s_n/n})$, it can naturally be expected that the oracle estimator also has rate $a_n\asymp\sqrt{s_n/n}$, and as proved in Theorem \ref{thm:1} we can expect $b_n\asymp\sqrt{s_n\log(d\vee n)/n}$. Furthermore, Lemma \ref{lem:nablahora} provides an upper bound for $\|\nabla h_\gamma(\wh\vbeta^{ora})\|_\infty$. On the other hand, $\|\nabla h_\gamma^p(\wh\vbeta^{ora})\|_\infty$ depends on the estimation accuracy of $\wh\vbeta^p$ for $\vbeta_0$.
    \end{rmk}

    \begin{lem}\label{lem:nablahora}
		Assuming that conditions (A1)-(A3) hold and that $\|\wh\vbeta^{ora}-\vbeta_0\|=O_{p}(a_n)$, then with probability at least $1-(d \vee n)^{-C}$, we have:
		\bse
		\|\nabla h_\gamma(\wh\vbeta^{ora})\|_\infty=O_{p} \left(\gamma^2+a_n+\sqrt{\frac{a_n s_n\log(d \vee n)}{n}}+\frac{s_n\log^{3/2}(d \vee n)}{n}\right).
		\ese
    \end{lem}
The detailed proofs of the main theorems and lemmas are provided in the online supplementary materials.

\section{Discussion}

The escalating prevalence of obesity remains a severe threat to global public health, prompting urgent calls for innovative research and targeted interventions. Addressing the clinical and economic burdens requires effective analytical tools capable of disentangling the complex genetic architecture underlying obesity.

In this work, we developed and applied the KIQR framework to 
simultaneously select and estimate important SNPs associated with the 
upper quantiles of BMI, while incorporating valuable prior knowledge from large-scale meta-analyses. Traditional GWAS, despite its widespread use, 
faces severe power limitations in cohort studies such as the FHS, 
with a limited number of participants but over 500,000 SNPs. KIQR moves beyond marginal analyses by jointly modeling the upper conditional quantiles of BMI most relevant to clinical obesity.

We discovered several novel and interesting SNPs plausibly associated with obesity at a high quantile, rs3798696 in \textit{TFAP2A} (regulating lipid accumulation), rs7070523 in \textit{ITIH5} (linked to adipose inflammation), and rs178260 in \textit{AIFM3} (implicated in extreme obesity), which may provide new directions for future research. 

Many of the identified SNPs can also be validated by the literature as well as lab-based biological research. We further demonstrate the performance of our proposed Knowledge Integration Quantile Regression (KIQR) approach through numerical simulations. In our simulation mimicking real genetic data, the KIQR method utilizing correct prior information outperforms the traditional penalized quantile regression and quantile-regression-extended GWAS, while showing robustness to potential misspecification. 

This study contributes directly to personalized obesity treatment strategies by enabling clinicians to identify genetic variants associated with high BMI quantiles rather than relying solely on mean-based associations. More broadly, the KIQR framework, 
complemented by our computationally efficient algorithms, serves as 
a flexible analytical tool for exploring other complex traits where 
clinical interest lies in specific quantiles of the phenotypic 
distribution, including elevated blood pressure, abnormal glucose 
levels, and hypercholesterolemia.

\bibliographystyle{asa}		
\bibliography{essay1} 

\end{document}





\renewcommand{\thepage}{}

\begin{singlespace}
\title{ \bf Identifying Obesity Genetic Variants in Ultra-High-Dimension: A Knowledge Integration Quantile Regression Approach \vspace{-1em}}
\date{}
\maketitle

\pagestyle{plain}
\pagenumbering{arabic}

\appendix
\begin{center}
{\LARGE {\sc Supplementary Materials}}
\end{center}
\end{singlespace}

In the supplementary materials, we provide detailed proofs of the theorems and lemmas in Section \ref{sec:sketchproof}. In Section~\ref{sec:theoretical_properties}, we reproduce the Theoretical Properties in the main paper to enhance readability.

\numberwithin{equation}{section}
\numberwithin{table}{section}
\numberwithin{figure}{section}

\section{Theoretical Properties}
\label{sec:theoretical_properties}
    As in the main paper, we define our Knowledge Integration Quantile Regression (KIQR) estimator as: 
    \be
	\wh\vbeta^{KIQR}=\argmin_{\vbeta}(1-\zeta)h_\gamma(\vbeta)+\zeta h_\gamma^p(\vbeta)+ (1 - \zeta)\sum_{j=1}^{d_n}p'_{\lambda}(| \hat{\beta}^{(0)}_j |)|\beta_j|.
	\label{eq:KIQR}
	\ee
    Theoretically, we establish the oracle properties for the estimator from our Knowledge Integration Quantile Regression (KIQR) method with Huber loss approximation for the quantile check loss and local linear approximation of the SCAD penalty. Again, we omit subscript $\tau$ for different conditional quantiles throughout. 

    We first establish an oracle inequality for the KIQR estimator with a Lasso penalty or $L_1$ penalty
	\be
	\wh\vbeta^{KIQR0}=\argmin_{\vbeta}(1-\zeta)h_\gamma(\vbeta)+\zeta h_\gamma^p(\vbeta)+(1-\zeta)\lambda\|\vbeta\|_1.
	\label{eq:KIQR0}
	\ee
    Then, starting from the KIQR estimator with Lasso penalty, we show that the KIQR estimator with SCAD penalty $\hat{\boldsymbol{\beta}}^{KIQR}$ defined in equation (\ref{eq:KIQR}) has the appealing oracle property. Note that the values of $\lambda$ in the above two equations may be different. In the following, $C$ denotes a generic positive constant which may assume different values at different places. The following assumptions are imposed.
	\begin{itemize}
		\item[(A1)] $\vx_i$ is sub-Gaussian in the sense that $E\exp\{\vx_i\trans\va\}\le C\exp\{C\|\va\|^2\}$ for any vector $\va$. The matrix $\vSigma=E[\vx_i\vx_i\trans]$ has eigenvalues bounded and bounded away from zero.
		\item[(A2)] The conditional density of $\ep_i=y_i-\vx_i\trans\vbeta_0$, $f_\ep(.|\vx)$, is uniformly bounded by a constant $\bar f$, and its derivative is uniformly bounded by another constant $\bar{f'}$. On any interval $[-C,C]$, $f_\ep(.|\vx)$ is bounded below by a constant $\underline{f}>0$ ($\underline{f}$ depends on $C$ but this dependence is suppressed in the notation for simplicity).
		\item[(A3)] The true parameter $\vbeta_0$ is sparse with support $S=\{j:\beta_{0j}\neq 0\}$ and $|S|=s_n$. 
	\end{itemize}
	
	\begin{thm}\label{thm:1_app}
		Under assumptions (A1)-(A3), also assuming that $\|  \hat{\vbeta}^p-\vbeta_0\|\le c\gamma $ for a sufficiently small constant $c$, with $\lambda\ge 2(1-\zeta)\|\nabla h_\gamma(\vbeta_0)\|_\infty+2\zeta\|\nabla h_\gamma^p(\vbeta_0)\|_\infty$,  $\gamma>>\sqrt{s_n\log(d_n \vee n)/n}$, $\gamma>>\lambda\sqrt{s_n}$, we have with probability at least $1-(d_n \vee n)^{-C}$ that the estimator of equation (\ref{eq:KIQR0}) satisfies the following oracle inequality
		\bse
		\|\wh\vbeta^{KIQR0}-\vbeta_0\|\le C\lambda\sqrt{s_n}\; \mbox{ and } \|\wh\vbeta^{KIQR0}-\vbeta_0\|_1\le C\lambda s_n.
		\ese
	\end{thm}
	
	\begin{rmk}
		By Theorem \ref{thm:1_app}, we can take $\lambda=C(1-\zeta)\sqrt{\frac{\log(d_n\vee n)}{n}}+2\zeta\|\nabla h_\gamma^p(\vbeta_0)\|_\infty$. For the standard LASSO penalized quantile regression without the $h_\gamma^p$ term for prior knowledge, we would take $\lambda=C\sqrt{\frac{\log(d_n \vee n)}{n}}$ which would also lead to $\|\wh\vbeta^{KIQR0}-\vbeta_0\|\le C\lambda\sqrt{s_n}$ (here $\wh\vbeta^{KIQR0}=\wh\vbeta^{qLASSO}$ is the quantile LASSO estimator without using $h_\gamma^p$, equivalent to setting $\zeta=0$). Thus if $\|\nabla h_\gamma^p(\vbeta_0)\|_\infty$ is of a smaller order than $\sqrt{\frac{\log(d_n \vee n)}{n}}$, the Knowledge Integration Quantile Regression (KIQR) estimator uses a smaller $\lambda$ and thus the upper bound is smaller. This observation is similar to that made in \cite{Jiang2016}.
	\end{rmk}

    The proof of Theorem \ref{thm:1_app} utilizes the following lemmas.
	
	\begin{lem}\label{lem:convex_app}  Let $c>0$ be a sufficiently small constant. With the same assumptions used in Theorem \ref{thm:1_app},
		uniformly over the set $\Omega:=\{(\vbeta_1,\vbeta_2): \|\vbeta_1-\vbeta_2\|\le c\gamma, \|\vbeta_2-\vbeta_0\|\le c, \|\vbeta_2-\vbeta_0\|_1\le 3\sqrt{s_n}\|\vbeta_2-\vbeta_0\|\}$, we have with probability at least $1-e^{-t}$,
		\bse
		&&h_\gamma( \vbeta_1)-  h_\gamma( \vbeta_2)- \langle \nabla h_\gamma( \vbeta_2),\vbeta_1-\vbeta_2\rangle\\
		&\ge& \left(\frac{\underline{f}\lambda_{\min}(\vSigma)}{4}    - \frac{C\sqrt{s_n}}{\gamma}\left( \sqrt{\frac{\gamma\log (d_n)}{n}}+\frac{\log (d_n)}{n} \right)-C\sqrt{\frac{t}{n\gamma}}-C\frac{t}{n\gamma}  \right) \|\vbeta_1-\vbeta_2\|^2,
		\ese
		where $\lambda_{\min}(\vSigma)$ is the minimum eigenvalue of $\vSigma=E[\vx_i\vx_i\trans]$.
	\end{lem}

	\begin{lem}\label{lem:convex2_app} 
		Assume $c$ is a sufficiently small positive constant. With the same assumptions used in Theorem \ref{thm:1_app}, we have that, with probability at least $1-e^{-t}$, uniformly over the set $\Omega'=\{(\vbeta_1,\vbeta_2):\|\vbeta^p-\vbeta_2\|\le c\gamma, \|\vbeta_1-\vbeta_2\|\le c\gamma,\|\vbeta_1-\vbeta_2\|_1\le 3\sqrt{s_n}\|\vbeta_1-\vbeta_2\| \}$, we have
		\bse
		h_\gamma^p( \vbeta_1)- h_\gamma^p(\vbeta_2)-  \langle \nabla h_\gamma^p(\vbeta_2),\vbeta_1-\vbeta_2\rangle\ge \left(\frac{\lambda_{\min}(\vSigma)}{2}-C\left(\frac{1}{\gamma}  \sqrt{\frac{ s_n\log (d_n)}{n}}   +\sqrt{\frac{t}{n\gamma^2}}+\frac{t}{n\gamma}\right)    \right)\|\vbeta_1-\vbeta_2\|^2.
		\ese
	\end{lem}
	
	\begin{lem}\label{lem:nablah_app}
		With probability at least $1-e^{-t}$,
		\bse
		\|\nabla h_\gamma(\vbeta_0)\|_\infty\le C\left(\gamma^2+\sqrt{\frac{t+\log (d_n)}{n}}+\frac{t+\log (d_n)}{n}\right).
		\ese
	\end{lem}
	
	We now establish the oracle property for our Knowledge Integration Quantile Regression (KIQR) estimator with Huber loss approximation of the quantile check loss and local linear approximation of the SCAD penalty.

    Define $\wh\vbeta^{ora}=(\wh\vbeta^{ora}_S,\vnull)=\argmin_{\vbeta:\vbeta_{S^c}=\vnull}(1-\zeta)h_\gamma(\vbeta)+\zeta h_\gamma^p(\vbeta)$. Theorem \ref{thm:2_app} shows the oracle property that, $\hat{\boldsymbol{\beta}}^{KIQR}$ defined in equation (\ref{eq:KIQR}), using $\wh\vbeta^{KIQR0}$ as the initial estimator, is equal to $\wh\vbeta^{ora}$ with probability approaching one.

	\begin{thm}\label{thm:2_app}
		Assume $\|\wh\vbeta^{ora}-\vbeta_0\|\le a_n$ and $\|\wh \vbeta^{KIQR0} -\vbeta_0\|\le b_n$ for some positive sequences $a_n,b_n=o(1)$. If $b_n<<\lambda<<\min_{j\in S}|\beta_{0j}|$ and $\lambda>(1-\zeta)\|\nabla_\gamma(\hat\vbeta^{ora})\|_\infty+\zeta\|\nabla h_\gamma^p(\wh\vbeta^{ora})\|_\infty$, for sufficiently large  $n$, we have that the KIQR estimator of equation (\ref{eq:KIQR}) is the oracle estimator
		$$\wh\vbeta^{KIQR}=\wh\vbeta^{ora}.$$
	\end{thm}
	\begin{rmk} In the statement of Theorem \ref{thm:2_app}, we condition on $\|\wh\vbeta^{ora}-\vbeta_0\|\le a_n$, $\|\wh\vbeta^{KIQR0}-\vbeta_0\|\le b_n$ and the size of $(1-\zeta)\|\nabla h_\gamma(\wh\vbeta^{ora})\|_\infty+\zeta\|\nabla h_\gamma^p(\wh\vbeta^{ora})\|_\infty$, thus the statement is entirely deterministic.
		The rate of $\|\wh\vbeta^{ora}-\vbeta_0\|$ of course depends on that of $\|\wh\vbeta^p-\vbeta_0\|$. If $\wh\vbeta^p$ is far away from the true $\vbeta_0$, we cannot expect $\wh\vbeta^{ora}$ to be a good estimator unless $\zeta=0$. On the other hand, if $\|\wh\vbeta^p-\vbeta_0\|=O_p(\sqrt{s_n/n})$, it can naturally be expected that the oracle estimator also has rate $a_n\asymp\sqrt{s_n/n}$, and as proved in Theorem \ref{thm:1_app} we can expect $b_n\asymp\sqrt{s_n\log(d_n\vee n)/n}$. Furthermore, Lemma \ref{lem:nablahora_app} gives a bound for $\|\nabla h_\gamma(\wh\vbeta^{ora})\|_\infty$. On the other hand, $\|\nabla h_\gamma^p(\wh\vbeta^{ora})\|_\infty$ depends on how good $\wh\vbeta^p$ is as an estimator of $\vbeta_0$.
	\end{rmk}
	
	\begin{lem}\label{lem:nablahora_app}
		Assume assumptions (A1)-(A3) hold and that $\|\wh\vbeta^{ora}-\vbeta_0\|=O_{p}(a_n)$, then with probability at least $1-(d_n \vee n)^{-C}$, 
		\bse
		\|\nabla h_\gamma(\wh\vbeta^{ora})\|_\infty=O_{p} \left(\gamma^2+a_n+\sqrt{\frac{a_n s_n\log(d_n \vee n)}{n}}+\frac{s_n\log^{3/2}(d_n \vee n)}{n}\right).
		\ese
	\end{lem}
	
\section{Theoretical Proof}
For simplicity in notation, the following proof omits the intercept term. Including the intercept leads to only minor differences in the proof, though it complicates the notation considerably.
\label{sec:sketchproof}
\subsection{Proof of Theorem 1}
 
	\noindent\textbf{Proof of Theorem \ref{thm:1_app}.} For simplicity of notation, we write $\wh\vbeta^{KIQR0}$ as $\wh\vbeta$ in the current proof.  
	We have the basic inequality
	\be\label{eqn:basic}
	(1-\zeta)h_\gamma(\wh\vbeta)+\zeta h_\gamma^p(\wh\vbeta)+(1 - \zeta)\lambda\|\wh\vbeta\|_1\le(1-\zeta)h_\gamma(\vbeta_0)+\zeta h_\gamma^p(\vbeta_0)+(1 - \zeta)\lambda\|\vbeta_0\|_1.
	\ee
	Using the convexity of $h_\gamma(\vbeta)$ and $h_\gamma^p(\vbeta)$, we have $(1-\zeta)h_\gamma(\wh\vbeta)+\zeta h_\gamma^p(\wh\vbeta)-(1-\zeta)h_\gamma(\vbeta_0)-\zeta h_\gamma^p(\vbeta_0)\ge \langle (1-\zeta)\nabla h_\gamma(\vbeta_0)+\zeta\nabla h_\gamma^p(\vbeta_0),\wh\vbeta-\vbeta_0\rangle$, and we get
	\bse
	\langle (1-\zeta)\nabla h_\gamma(\vbeta_0)+\zeta\nabla h_\gamma^p(\vbeta_0),\wh\vbeta-\vbeta_0\rangle\le  (1 - \zeta)(\lambda\|\vbeta_0\|_1-\lambda\|\wh\vbeta\|_1).
	\ese
	Using $\|(1-\zeta)\nabla h_\gamma(\vbeta_0)+\zeta\nabla h_\gamma^p(\vbeta_0)\|_\infty\le (1 - \zeta)\lambda/2$, we get
	\bse
	0&\le &\frac{\lambda}{2}\|\wh\vbeta-\vbeta_0\|_1+\lambda\|\vbeta_0\|_1-\lambda\|\wh\vbeta\|_1\\
	&=& \frac{\lambda}{2}\|(\wh\vbeta-\vbeta_0)_S\|_1+\frac{\lambda}{2}\|\wh\vbeta_{S^c}\|_1+\lambda\|(\vbeta_0)_S\|_1-\lambda\|\wh\vbeta_{S}\|_1-\lambda\|\wh\vbeta_{S^c}\|_1\\
	&\le& \frac{\lambda}{2}\|(\wh\vbeta-\vbeta_0)_S\|_1+\frac{\lambda}{2}\|\wh\vbeta_{S^c}\|_1+\lambda\|(\wh\vbeta-\vbeta_0)_S\|_1-\lambda\|\wh\vbeta_{S^c}\|_1.
	\ese
	Thus
	\be\label{eqn:cone}
	\|(\wh\vbeta-\vbeta_0)_{S^c}\|_1\le 3\|(\wh\vbeta-\vbeta_0)_{S}\|_1.
	\ee
	If $\|\wh\vbeta-\vbeta_0\|\le c\gamma$ for a sufficiently small $c$, using Lemmas \ref{lem:convex_app} and \ref{lem:convex2_app} (taking $\vbeta_2=\vbeta_0$ in these two lemmas), we then have
	\bse
	C\|\wh\vbeta-\vbeta_0\|^2&\le& \langle (1-\zeta)\nabla h_\gamma(\vbeta_0)+\zeta\nabla h_\gamma^p(\vbeta_0),\wh\vbeta-\vbeta_0\rangle+(1-\zeta)(\lambda\|\vbeta_0\|_1-\lambda\|\wh\vbeta\|_1)\\
	&\le&\frac{\lambda}{2}\|(\wh\vbeta-\vbeta_0)_S\|_1+\frac{\lambda}{2}\|\wh\vbeta_{S^c}\|_1+\lambda\|(\wh\vbeta-\vbeta_0)_S\|_1-\lambda\|\wh\vbeta_{S^c}\|_1\\
	&\le&\frac{3\lambda}{2}\|(\wh\vbeta-\vbeta_0)_S\|_1\le \frac{3\lambda\sqrt{s}}{2}\|\wh\vbeta-\vbeta_0\|.
	\ese
	This means $\|\wh\vbeta-\vbeta_0\|\le C\lambda\sqrt{s}$ and $\|\wh\vbeta-\vbeta_0\|_1\le C\lambda s$ follows immediately from \eqref{eqn:cone}.
	
	On the other hand, if $\|\wh\vbeta-\vbeta_0\|> c\gamma$. Define $\wt\vbeta=(1-\alpha)\vbeta_0+\alpha\wh\vbeta$ with $\alpha=\frac{c\gamma}{\|\wh\vbeta-\vbeta_0\|}\in (0,1)$. By the definition of $\wt\vbeta$, we easily see 
	\be\label{eqn:equalcg}
	\|\wt\vbeta-\vbeta_0\|=c\gamma.
	\ee
	By the convexity of $h_\gamma(.)$, we have
	\be\label{eqn:basic2}
	(1-\zeta)h_\gamma(\wt\vbeta)+\zeta h_\gamma^p(\wt\vbeta)+\lambda\|\wt\vbeta\|_1\le(1-\zeta)h_\gamma(\vbeta_0)+\zeta h_\gamma^p(\vbeta_0)+(1 - \zeta)\lambda\|\vbeta_0\|_1.
	\ee
	With \eqref{eqn:basic2} in place of \eqref{eqn:basic}, the same arguments as above lead to $\|\wt\vbeta-\vbeta_0\|\le C\lambda\sqrt{s}=o(\gamma)$, which is a contradiction to \eqref{eqn:equalcg}. This completes the proof of the theorem.
	\hfill $\Box$
	
	Lemmas 1 and 2 are used in the proof of Theorem \ref{thm:1_app}, establishing the local strong convexity of the loss function. Lemma \ref{lem:nablah_app} further establishes a bound for $\|\nabla h_\gamma(\vbeta^0)\|_\infty$ which is related to the choice of $\lambda$ in the lasso estimator.
	
	\noindent\textbf{Proof of Lemma \ref{lem:convex_app}.}
	Define the events $E_{i1}=\{|\langle \vx_i,\frac{\vbeta_1-\vbeta_2}{\|\vbeta_1-\vbeta_2\|}\rangle|\le \frac{ 1}{2c}\}$, $E_{i2}=\{|y_i-\vx_i\trans\vbeta_0|\le \frac{\gamma}{2}\}=\{|\ep_i+\langle \vx_i,\vbeta_0-\vbeta_2\rangle|\le \frac{\gamma}{2}\}$, $E_{i3}=\{|\langle \vx_i,\vbeta_0-\vbeta_2\rangle|\le C\}$, $E_i=E_{i1}\cup E_{i2}\cup E_{i3}$. 
	
  Using $P(E_{i2}E_{i3}|\vx_i)\ge P(\ep_i\in[\langle\vx_i,\vbeta_2-\vbeta_0\rangle-\frac{\gamma}{2}, \langle\vx_i,\vbeta_2-\vbeta_0\rangle+\frac{\gamma}{2}]|\vx_i)I_{E_{i3}}\ge I_{E_{i3}}\underline{f} \gamma$, we have
	\bse
	&&E[\langle \vx_i,\vbeta_1-\vbeta_2\rangle^2I_{E_i}]\\
	&\ge & \underline{f}\gamma E[\langle \vx_i,\vbeta_1-\vbeta_2\rangle^2I_{E_{i1}\cap E_{i3}} ]\\
	&\ge &\underline{f}\gamma\left(E[\langle \vx_i,\vbeta_1-\vbeta_2\rangle^2]-E[\langle \vx_i,\vbeta_1-\vbeta_2\rangle^2I_{E_{i1}^c} ]-E[\langle \vx_i,\vbeta_1-\vbeta_2\rangle^2I_{E_{i3}^c}]\right)\\
	&\ge &\frac{\underline{f}\gamma}{2} E[\langle \vx_i,\vbeta_1-\vbeta_2\rangle^2,
	\ese
	where in the last step we used that 
	\bse
	&&E[\langle \vx_i,\vbeta_1-\vbeta_2\rangle^2I_{E_{i3}^c}]\\
	&\le & (E\langle \vx_i,\vbeta_1-\vbeta_2\rangle^4)^{1/2} (P(|\langle \vx_i,\vbeta_0-\vbeta_2\rangle|> C))^{1/2}\\
	&\le &CE[\langle \vx_i,\vbeta_1-\vbeta_2\rangle^2]\exp\{-\frac{C}{\|\vbeta_0-\vbeta_2\|^2}\}\\
	&\le&\frac{1}{4}E[\langle \vx_i,\vbeta_1-\vbeta_2\rangle^2],
	\ese
	when $c$ is sufficiently small, and that similarly,
	\bse
	&& E[\langle \vx_i,\vbeta_1-\vbeta_2\rangle^2I_{E_{i1}^c}]\\
	&\le &(E\langle \vx_i,\vbeta_1-\vbeta_2\rangle^4)^{1/2} \left(P(|\langle \vx_i,\vbeta_1-\vbeta_2\rangle|> \frac{\|\vbeta_1-\vbeta_2\|}{2c}\right)^{1/2}\\
	&\le &CE[\langle \vx_i,\vbeta_1-\vbeta_2\rangle^2]\exp\{-\frac{C}{c^2}\}\\
	&\le&\frac{1}{4}E[\langle \vx_i,\vbeta_1-\vbeta_2\rangle^2].
	\ese
	
	Define $g(\vx_i;\vdelta)=\frac{1}{\gamma}(\vx_i\trans\vdelta)^2 I_{E_i}$, with $\vdelta=\vbeta_1-\vbeta_2/\|\vbeta_1-\vbeta_2\|$. Note we have $\|\vdelta\|_1\le 3\sqrt{s_n}$ by the definition of $\Omega$. We have
	\bse
	&&E\left[\sup_{\Omega}|(P-P_n)g(\vx_i;\vdelta)|\right]\\
	&\le& 2E\left[\sup_{\Omega}\left|\frac{1}{n}\sum_i\sigma_i g(\vx_i;\vdelta)\right|\right]\\
	&=&2E\left[\sup_{\Omega}\left|\frac{1}{n}\sum_i\sigma_i a_i(\vx_i\trans\vdelta_i)I_{E_i}\right|\right]\\
	&\le&\frac{C}{\gamma}E\left[\sup_{\Omega}\left|\frac{1}{n}\sum_i\sigma_i  (\vx_i\trans\vdelta)I_{E_i}\right|\right]\\
	&\le &\frac{C}{\gamma} E\left[\sup_{\Omega}\left\|\frac{1}{n}\sum_i\sigma_i \vx_i I_{E_i}\right\|_\infty\|\vdelta\|_1\right],
	\ese
	where the first step used the symmetrization technique (Theorem 2.1 of \cite{koltchinskiibook2011}) with $\sigma_i\in\{-1,1\}$ being the binary Rademacher variables, in the second step we defined $a_i=\frac{1}{\gamma}\vx_i\trans\vdelta I_{E_i}$ with $|a_i|\le \frac{C}{\gamma}$, and the third step used the contraction inequality for the Rademacher processes (Theorem 4.4 of \cite{ledoux91}).
	
	Using the sub-Gaussianty of $\vx_i$ and that $E[|x_{ij}I_{E_i}|^k]\le k!\gamma C^{k-2}, k\ge 2$, we have by Bernstein's inequality and taking union over $j\in\{1,\ldots,p\}$,  
	\bse
	P(\|\frac{1}{n}\sum_i\sigma_i\vx_{i}I_{E_i}\|_\infty>t)\le p\exp\{-Cn\frac{t^2}{t+\gamma}\},
	\ese
	and the above is equivalent to
	\bse
	P(\|\frac{1}{n}\sum_i\sigma_i\vx_{i}I_{E_i}\|_\infty>a(t))\le e^{-t},
	\ese
	with $a(t)=C\left(\sqrt{t}\sqrt{\frac{\gamma}{n}}+\frac{t}{n}+\sqrt{\frac{\gamma\log d_n}{n}}+\frac{\log d_n}{n}\right)$.
	Denoting $X=\|\frac{1}{n}\sum_i\sigma_i\vx_{i}I_{E_i}\|_\infty$ and using $E[X]\le \sum_{t=0}^\infty EI\{a(t-1)\le X\le a(t)\}a(t)\le \sum_{t=0}^\infty Ce^{-t}a(t)$, with the convention $a(-1)=0$, we get
	\bse
	E\left[\left\|\frac{1}{n}\sum_i\sigma_i\vx_{i}I_{E_i}\right\|_\infty\right]\le C\left( \sqrt{\frac{\gamma\log d_n}{n}}+\frac{\log d_n}{n}\right).
	\ese
	Using Talagrand's concentration equality, since    $|g(\vx_i;\vdelta)|\le\frac{C}{\gamma}$ and $E[g^2(\vx_i;\vdelta)]\le E[\frac{C\bar f}{\gamma}(\vx_i\trans\vdelta)^4 I_{E_{i1}}]\le \frac{C }{\gamma},$ then with probability $1-e^{-t}$,
	\bse
	|\sup_{\Omega}(P-P_n)g(\vx_i;\vdelta)|&\le& C\left(E[\sup_{\Omega}(P-P_n)g(\vx_i;\vdelta)]+\sqrt{\frac{t}{n\gamma}}+\frac{t}{n\gamma}\right)\\
	&\le&C\left(\frac{\sqrt{s_n}}{\gamma}\left( \sqrt{\frac{\gamma\log d_n}{n}}+\frac{\log d_n}{n} \right)+\sqrt{\frac{t}{n\gamma}}+\frac{t}{n\gamma}\right).
	\ese
	
	Thus 
	\bse
	P_ng(\vx_i;\vdelta)&\ge& Eg(\vx_i;\vdelta)]-C\left(\frac{\sqrt{s_n}}{\gamma}\left( \sqrt{\frac{\gamma\log d_n}{n}}+\frac{\log d_n}{n} \right)+\sqrt{\frac{t}{n\gamma}}+\frac{t}{n\gamma}\right)\\
	&\ge& \frac{\underline{f}}{2}E[(\vx_i\trans\vdelta)^2]-C\left(\frac{\sqrt{s_n} }{\gamma}\left( \sqrt{\frac{\gamma\log d_n}{n}}+\frac{\log d_n}{n} \right)+\sqrt{\frac{t}{n\gamma}}+\frac{t}{n\gamma}\right),
	\ese
	or equivalently,
	\be\label{eqn:convexinner2}
	&&\frac{1}{n}\sum_i \frac{1}{\gamma}\langle\vx_i,\vbeta_1-\vbeta_2\rangle^2 I_{E_i}\nonumber\\
	&\ge& \left(\frac{\underline{f}\lambda_{\min}(\vSigma)}{2}    - \frac{C\sqrt{s_n}}{\gamma}\left( \sqrt{\frac{\gamma\log d_n}{n}}+\frac{\log d_n}{n} \right)-C\sqrt{\frac{t}{n\gamma}}-C\frac{t}{n\gamma}  \right) \|\vbeta_1-\vbeta_2\|^2.
	\ee
	
	When $|y_i-\vx_i\trans\vbeta_1|\le \gamma$ and $|y_i-\vx_i\trans\vbeta_2|\le \gamma$, which is implied by $E_i$, we have  $\nabla h_\gamma(y_i-\vx_i\trans\vbeta_1)- \nabla h_\gamma(y_i-\vx_i\trans\vbeta_2)=\frac{1}{2\gamma}\langle\vx_i,\vbeta_1-\vbeta_2\rangle^2$ and thus
	\bse
	&&  h_\gamma(  \vbeta_1)-  h_\gamma( \vbeta_2)- \langle \nabla h_\gamma( \vbeta_2),\vbeta_1-\vbeta_2\rangle\\
	&=&\frac{1}{2n\gamma}\sum_i\langle\vx_i,\vbeta_1-\vbeta_2\rangle^2I_{E_i}\\
	&\ge&\left(\frac{\underline{f}\lambda_{\min}(\vSigma)}{4}    - \frac{C}{\gamma}\left( \sqrt{\frac{\gamma\log d_n}{n}}+\frac{\log d_n}{n} \right)-C\sqrt{\frac{t}{n\gamma}}-C\frac{t}{n\gamma}  \right) \|\vbeta_1-\vbeta_2\|^2.
	\ese
	\hfill $\Box$
	
	\noindent\textbf{Proof of Lemma \ref{lem:convex2_app}.} The proof is similar to that of Lemma \ref{lem:convex_app}.
	Denote $\vdelta=(\vbeta_1-\vbeta_2)/\|\vbeta_1-\vbeta_2\|$
Let $E_{i1}=\{|\langle\vx_i,\vdelta\rangle|\le \frac{1}{2c}\}$, $E_{i2}=\{|\langle\vx_i,\vbeta^p-\vbeta_2\rangle|\le \gamma/2\}$, $E_i=E_{i1}\cup E_{i2}$.
	We have
	\bse
	&&E[\langle\vx_i,\vbeta_1-\vbeta_2\rangle^2I_{E_i}]\\
	&\ge &E[\langle\vx_i,\vbeta_1-\vbeta_2\rangle^2]-E[\langle\vx_i,\vbeta_1-\vbeta_2\rangle^2I_{E_{i1}^c}]-E[\langle\vx_i,\vbeta_1-\vbeta_2\rangle^2I_{E_{i2}^c}]\\
	&\ge& \gamma E[\langle\vx_i,\vbeta_1-\vbeta_2\rangle^2],
	\ese
	due to that
	\bse
	&&E[\langle\vx_i,\vbeta_1-\vbeta_2\rangle^2I_{E_{i2}^c}]\\
	&\le &(E[\langle\vx_i,\vbeta_1-\vbeta_2\rangle^4)^{1/2}P(E_{i2}^c)^{1/2}\\
	&\le &CE[\langle\vx_i,\vbeta_1-\vbeta_2\rangle^2]\exp\{-\frac{C\gamma^2}{\|\vbeta^p-\vbeta_2\|^2}\}\\
	&\le &\frac{1-\gamma}{2}E[\langle\vx_i,\vbeta_1-\vbeta_2\rangle^2],
	\ese
	and
	\bse
	&&E[\langle\vx_i,\vbeta_1-\vbeta_2\rangle^2I_{E_{i1}^c}]\\
	&\le &(E[\langle\vx_i,\vbeta_1-\vbeta_2\rangle^4)^{1/2}P(E_{i1}^c)^{1/2}\\
	&\le &CE[\langle\vx_i,\vbeta_1-\vbeta_2\rangle^2]\exp\{-\frac{C}{c^2}\}\\
	&\le &\frac{1-\gamma}{2}E[\langle\vx_i,\vbeta_1-\vbeta_2\rangle^2].
	\ese
	
	Let $g(\vx_i;\vdelta)=\frac{1}{\gamma}(\vx_i\trans\vdelta)^2I_{E_i}$. We have
	\bse
	&&E\left[\sup_{\Omega'}|(P-P_n)g(\vx_i;\vdelta)|\right]\\
	&\le &\frac{C}{\gamma} E\left[\sup_{\Omega'}\left\|\frac{1}{n}\sum_i\sigma_i \vx_i\right\|_\infty\|\vdelta\|_1\right]\\
	&\le &\frac{C}{\gamma}\sqrt{\frac{s_n\log d_n}{n}}.
	\ese

	Using that $|g(\vx_i;\vdelta)|\le\frac{C}{\gamma}$, the concentration inequality implies that with probability $1-e^{-t}$,
	\bse
	|\sup_{\Omega'}(P-P_n)g(\vx_i;\vdelta)|
	&\le&C\left(\frac{1}{\gamma}  \sqrt{\frac{s_n \log d_n}{n}}   +\sqrt{\frac{t}{n\gamma^2}}+\frac{t}{n\gamma}\right).
	\ese
	
	Thus  
	\bse
	&&\frac{1}{n}\sum_i \frac{1}{\gamma}\langle\vx_i,\vbeta_1-\vbeta_2\rangle^2 I_{E_i}\\
	&\ge& \left(\lambda_{\min}(\vSigma)-C\left(\frac{1}{\gamma}  \sqrt{\frac{s_n \log d_n}{n}}   +\sqrt{\frac{t}{n\gamma^2}}+\frac{t}{n\gamma}\right)    \right)\|\vbeta_1-\vbeta_2\|^2.
	\ese
	
	The rest of the proof is identical to the proof of Lemma \ref{lem:convex_app} and thus omitted.
	\hfill $\Box$
	
	\noindent\textbf{Proof of Lemma \ref{lem:nablah_app}.}	
	Let $H(u)=h_\gamma'(u)=\frac{u}{2\gamma}I\{|u|\le \gamma\}+\frac{1}{2}sign(u)I\{|u|>\gamma\}+\tau-\frac{1}{2}$. It can be verified $H(u)=\frac{1}{2}\int_{-1}^1(\tau-I\{u\le \gamma v\})dv$. 
	The $j$-th component of $\nabla h_\gamma(\vbeta_0)$ is $\nabla_j h_\gamma(\vbeta_0)=-(1/n)\sum_i x_{ij}H(y_i-\vx_i\trans\vbeta_0)$. Using
	\be\label{eqn:EH}
	\left|E[H(y_i-\vx_i\trans\vbeta_0)|\vx_i]\right|&=&\left|E[\frac{1}{2}\int_{-1}^1(\tau-I\{\ep_i\le \gamma v\})dv |\vx_i]\right|\nonumber\\
	&=&\left|\frac{1}{2}\int_{-1}^1 F_{\ep}(0|\vx_i)-F_{\ep}(\gamma v|\vx_i)dv\right|\nonumber\\
	&=&\left|\frac{1}{2}\int_{-1}^1 f_{\ep}(0|\vx_i)\gamma v+\frac{\bar{f'}}{2}\gamma^2v^2dv\right|\nonumber\\
	&\le &\frac{\bar{f'}\gamma^2}{6},
	\ee
	where $F_\ep$ is the conditional cumulative distribution function of $\ep$,
	we get $\|E\nabla h_\gamma(\vbeta_0)\|_\infty=O(\gamma ^2)$.
	
	Using $|H|\le 1$ and the Bernstein's inequality, with probability $1-d_ne^{-t}$, 
	\bse
	&&\|\nabla h_\gamma(\vbeta_0)-E[\nabla h_\gamma(\vbeta_0)]\|_\infty\le C\left(\sqrt{\frac{t}{n}}+\frac{t}{n}\right),
	\ese
	which completes the proof.
	\hfill $\Box$

 \subsection{Proof of Theorem 2}
    Define $\wh\vbeta^{ora}=(\wh\vbeta^{ora}_S,\vnull)=\argmin_{\vbeta:\vbeta_{S^c}=\vnull}(1-\zeta)h_\gamma(\vbeta)+\zeta h_\gamma^p(\vbeta)$.
	We now show the oracle property that, using $\wh\vbeta^{KIQR0}$ as the initial estimator, $\hat{\boldsymbol{\beta}}^{KIQR}$ defined in equation (\ref{eq:KIQR}) is equal to $\wh\vbeta^{ora}$ with probability approaching one.
 
	\noindent\textbf{Proof of Theorem \ref{thm:2_app}.} 
	We only need to prove that in a sufficiently small neighborhood $0<\|\vbeta-\wh\vbeta^{ora}\|\le c$ (with $c$ sufficiently small), 
	\label{eqn:diffloc}
    \begin{align}
	(1-\zeta)h_\gamma(\vbeta)+\zeta h_\gamma^p(\vbeta) &+(1 - \zeta)\sum_j\lambda_j|\beta_j|_1  \nonumber \\
    & >   (1-\zeta)h_\gamma(\wh\vbeta^{ora})+\zeta h_\gamma^p(\wh\vbeta^{ora})
    +(1 - \zeta)\sum_j\lambda_j|\wh\beta^{ora}_j|,
    \end{align}
	where $\lambda_j=p'_{\lambda}(|\wh\beta_j^{KIQR0}|)$. Since $|\wh\beta_j^{KIQR0}-\beta_{0j}|\le \|\wh\vbeta^{KIQR0}-\vbeta_0\|\le b_n<<\lambda$ and $\min_{j\in S}|\beta_{0j}|>>\lambda$, we have $\lambda_j=0$ for $j\in S$ and $\lambda_j=\lambda$ for $j\in S^c$ (since for the SCAD penalty, $p'_\lambda(x)=\lambda$ when $|x|\le \lambda$).
	Using the convexity of $h_\gamma(\vbeta)$,
	\bse
	&&(1-\zeta)h_\gamma(\vbeta)+\zeta h_\gamma^p(\vbeta)-(1-\zeta)h_\gamma(\wh\vbeta^{ora})-\zeta h_\gamma^p(\wh\vbeta^{ora})\\
	&\ge&\langle (1-\zeta)\nabla h_\gamma(\wh\vbeta^{ora})+\zeta\nabla h_\gamma^p(\wh\vbeta^{ora}),\vbeta-\wh\vbeta^{ora}\rangle\\
	&=&\sum_{j\in S^c}\left((1-\zeta)\nabla_j h_\gamma(\wh\vbeta^{ora})+\zeta\nabla_j h_\gamma^p(\wh\vbeta^{ora})\right) \beta_j,
	\ese
	where $\nabla_j h_\gamma(\wh\vbeta^{ora})$ is the $j$-th component of $\nabla h_\gamma(\wh\vbeta^{ora})$, and we used that $(1-\zeta)\nabla_j h_\gamma(\wh\vbeta^{ora})+\zeta\nabla_j h_\gamma^p(\wh\vbeta^{ora})=0, j\in S$ by the definition of $\wh\vbeta^{ora}$. Thus the difference of the two sides of \eqref{eqn:diffloc} is bounded below by
	\bse
	&&\sum_j (1-\zeta)\lambda_j|\beta_j|-\sum_j (1-\zeta)\lambda_j|\wh\beta^{ora}_j|+\sum_{j\in S^c}\left((1-\zeta)\nabla_j h_\gamma(\wh\vbeta^{ora})+\zeta\nabla_j h_\gamma^p(\wh\vbeta^{ora})\right) \beta_j\\
	&\ge&\sum_{j\in S^C}((1-\zeta)\lambda_j-\|(1-\zeta)\nabla h_\gamma(\wh\vbeta^{ora})+\zeta\nabla h_\gamma^p(\wh\vbeta^{ora})\|_\infty)|\beta_j|
	\ge 0,
	\ese
	where the last inequality used Lemma \ref{lem:nablahora_app} and the assumption on $\lambda$, and it is strictly positive unless $\beta_j=\wh\beta^{ora}_j,\forall j\in S^c$. By that $\wh\vbeta^{ora}$ is the minimizer with constraint that the support is contained in $S$, we have $\wh\vbeta^{KIQR}=\wh\vbeta^{ora}$.
	\hfill $\Box$
	
	\noindent\textbf{Proof of Lemma \ref{lem:nablahora_app}.}
	Similar to \eqref{eqn:EH}, we have
	\bse
	|E[x_{ij}H(y_i-\vx_i\trans\vbeta)|\vx_i]|&=&\left|\left(\frac{1}{2}\int_{-1}^1 f_{\ep}(0|\vx_i)(\gamma v+\vx_i\trans(\vbeta-\vbeta_0))+\frac{\bar{f'}}{2}(\gamma v+\vx_i\trans(\vbeta-\vbeta_0))^2dv\right) x_{ij}\right|\\
	&\le& C(\gamma^2+|\vx_i\trans(\vbeta-\vbeta_0)|+|\vx_i\trans(\vbeta-\vbeta_0)|^2)|x_{ij}|.
	\ese
	Thus 
	\be\label{eqn:comb1}
	\|E\nabla h_\gamma(\wh\vbeta^{ora})\|_\infty=O_p(\gamma^2+a_n).
	\ee
	We then note that
	\bse
	|x_{ij}\{H(y_i-\vx_i\trans\vbeta)-H(y_i-\vx_i\trans\vbeta_0)\}|
	& \le& |x_{ij}|,
	\ese
	and
	\bse
	&&E\left[x_{ij}^2\{H(y_i-\vx_i\trans\vbeta)-H(y_i-\vx_i\trans\vbeta_0)\}^2\right]\\
	&=&E\left[\frac{x_{ij}^2}{4}\left(\int_{-1}^1I\{\epsilon_i\le \gamma v\}-I\{\epsilon_i\le \gamma v+\vx_i\trans(\vbeta-\vbeta_0)\}dv\right)^2\right]\\
	&\le&E\left[\frac{x_{ij}^2}{2}\left|\int_{-1}^1I\{\epsilon_i\le \gamma v\}-I\{\epsilon_i\le \gamma v+\vx_i\trans(\vbeta-\vbeta_0)\}dv\right|\right]\\
	&\le&CE\left[ \left|\int_{-1}^1I\{\epsilon_i\le \gamma v\}-I\{\epsilon_i\le \gamma v+\vx_i\trans(\vbeta-\vbeta_0)\}dv\right|\right]\\
	&\le &CE|\vx_i\trans(\vbeta-\vbeta_0)|\le C\|\vbeta-\vbeta_0\|.
	\ese
	Define the class of functions $\calF_j=\{x_{j}I\{\|x\|_\infty\le c_n\} H(y-\vx\trans\vbeta): \|\vbeta-\vbeta\|\le a_n, \mbox{supp}\{\vbeta\} \in S\}$, where $c_n=C\sqrt{\log(d_n\vee n)}$. Based on Lemmas 2.6.15 and 2.6.18 in \cite{vaartwellner96}, $\calF_j$ is a VC-graph with VC-index bounded by $Cs$, and Theorem 2.6.7 there gives a covering number bound 
	\bse
	N(\delta,\calF_j, \|.\|_{L^2(P_n)})\le \left(\frac{C\|F_j\|_{L^2(P_n)}}{\delta}\right)^{Cs},
	\ese
	where $F_j=|x_j|$ is an envelope function for $\calF_j$.
	Thus, by Theorem 3.12 of \cite{koltchinskiibook2011}, 
	\bse
	E\left[\sup_{f\in\calF_j}\left| \frac{1}{n}\sum_i\sigma_if(\vx_i,y_i)\right|\right]\le C\left(\sqrt{\frac{a_ns_n\log(1/a_n)}{n}} + c_n\frac{s_n\log(1/a_n)}{n} \right).
	\ese
	
	By Talagrand's concentration inequality, with probability $1-e^{-t}$,
	\be\label{eqn:comb2}
	&&\sup_{f\in\calF_j}\left| \frac{1}{n}\sum_i\sigma_if(\vx_i,y_i)\right|\nonumber\\
	&\le&
	CE\left[\sup_{f\in\calF_j}\left| \frac{1}{n}\sum_i\sigma_if(\vx_i,y_i)\right|\right]+C\left(\sqrt{\frac{a_nt}{n}}+c_n\frac{t}{n}\right).
	\ee
	
	Finally, using symmetrization and union bound, with probability $1-pe^{-t}$,
	\be\label{eqn:comb3}
	&&\sup_{\|\vbeta-\vbeta_0\|\le a_n, \vbeta_{S^c}=\vnull}\|\nabla h_\gamma(\vbeta)-\nabla h_\gamma(\vbeta_0)-E\nabla h_\gamma(\vbeta)+E\nabla h_\gamma(\vbeta_0)\|_\infty\nonumber\\
	&\le&  C\left(\sqrt{\frac{a_ns\log(1/a_n)}{n}} + c_n\frac{s\log(1/a_n)}{n} \right)+C\left(\sqrt{\frac{a_nt}{n}}+c_n\frac{t}{n}\right).
	\ee
	The proof is completed by combining \eqref{eqn:comb1}, \eqref{eqn:comb2} and \eqref{eqn:comb3} (with $t\asymp \log(d_n\vee n)$).
	\hfill $\Box$
\bibliographystyle{asa}		
\bibliography{essay1}